\documentclass[aps,pra,reprint,amsmath,amssymb,groupedaddress,showpacs]{revtex4-1}
\usepackage{graphicx}%Include figure files
\usepackage{float}
\usepackage{dcolumn} %Align table columns on decimal point
\usepackage{bm}      %bold math
\usepackage[bookmarks=false]{hyperref} %Hyperlinks
\usepackage[font={small}]{caption} %FIXES FIGURE CAPTION PROBLEM
\hypersetup{
    unicode=false,          % non-Latin characters in Acrobat's bookmarks
    pdftoolbar=true,        % show Acrobat's toolbar?
    pdfmenubar=true,        % show Acrobat's menu?
    pdffitwindow=true,      % window fit to page when opened
    pdfstartview={},        % fits the width of the page to the window FitH
    pdftitle={Random-Unitary Depolarization Ensures the Correctablity of All Quantum Channels}, % title
    pdfauthor={Samuel R. Hedemann},% author
    pdfsubject={Quantum Error Correction},   % subject of the document
    pdfcreator={},  % creator of the document
    pdfproducer={}, % producer of the document
    pdfkeywords={Quantum,} {Error Correction,} {Random Unitary,} {Depolarization,} {Dephasing,} {Phase Damping,} {Quantum Operations,} {Kraus Operators,} {Shor Code}, % list of keywords
    pdfnewwindow=true,      % links in new window
    colorlinks=true,        % false: boxed links; true: colored links
    linkcolor=black,        % color of internal links (change box color with linkbordercolor)
    citecolor=black,        % color of links to bibliography
    filecolor=black,        % color of file links
    urlcolor=black          % color of external links
}%All links are clickable black text->least obtrusive, easy to find with mouse
\newcommand{\Sec}[1]{\hyperref[sec:#1]{Sec.{\kern 2pt}\ref*{sec:#1}}}
\newcommand{\Section}[1]{\hyperref[sec:#1]{Section~\ref*{sec:#1}}}
\newcommand{\Fig}[2][]{\hyperref[fig:#2]{Fig.{\kern 2pt}\ref*{fig:#2}#1}}%#1=a,b,etc. is the optional argument, in [] at call. Ex: \Fig[c]{2} = Fig. 2c
\newcommand{\Figure}[2][]{\hyperref[fig:#2]{Figure~\ref*{fig:#2}#1}}%#1=a,b,etc. is the optional argument, in [] at call. Ex: \Fig[c]{2} = Fig. 2c 
\begin{document}
%*******************************************************************************
%                                   TITLE
% \title{Random-Unitary Depolarization and the Reversibility of All Quantum Channels}
% \title{Random-Unitary Depolarization Allows the Correctablity of All Quantum Channels}
% \title{Hilbert-Schmidt Complete Random-Unitary Depolarization Ensures the Correctablity of All Quantum Channels}
\title{Random-Unitary Depolarization Ensures the Correctablity of All Quantum Channels}%RU Depolarization is sufficient because the depolarization requires the operators to span the space which gives them their HS completeness, and the RU property ensures their correctability.
%*******************************************************************************
%*******************************************************************************
%                                  BYLINES
\author{Samuel R. Hedemann}
\affiliation{Dept.~of Physics and Engineering Physics, Stevens Institute of Technology, Hoboken, NJ 07030, USA}
\date{\today}
%*******************************************************************************
%*******************************************************************************
\begin{abstract}%                 0. ABSTRACT
We prove that if any error channel has a Kraus decomposition that is simultaneously correctable and Hilbert-Schmidt (HS) complete, then the existence of Kraus sets with these properties guarantees the correctability of all quantum channels.  As a proof of the existence of such Kraus sets, the $n$-level depolarization channel is shown to have a random-unitary (RU) decomposition that is both HS complete and correctable due its RU nature, thereby proving that all quantum channels are correctable.  As an application, conditions for universal error-correction operations are presented.
\end{abstract}
%*******************************************************************************
%*******************************************************************************
%                             PACS & TITLE COMMAND
 %Quantum systems with finite Hilbert space
\maketitle
%*******************************************************************************
%*******************************************************************************
%                              I. INTRODUCTION
\section{\label{sec:I}Introduction}
As the demand for quantum computers increases, the need for reliable quantum error correction becomes more important, as well \cite{Feyn,DiV1}.  Many error-correction schemes have been proposed, such as the Shor code \cite{Shor}, the Steane code \cite{Stea}, and environment-assisted methods \cite{RU05}, which often employ random-unitary (RU) decompositions \cite{RU07,RU08,RU03,RU01,RU06,RU02,Xin1,RU04}, with a wide range of pioneering work in general quantum error correction found in \cite{BDSW,EkMa,LMPZ,KnLa,KnL2}.

Due to their inherent correctability, we focus on RU Kraus decompositions \cite{HKr1,HKr2,Choi,Kra1} of quantum channels for discrete systems.  While depolarization channels are well-known to have RU Kraus sets, we prove that these sets are \textit{also} Hilbert-Schmidt (HS) complete, and show that the existence of such correctable HS-complete Kraus sets guarantees the correctability of all quantum channels.

To review, a \textit{quantum operation} $\mathcal{E}$ is a map between physical states given by density operators $\rho$ and $\rho'$, as
%===============================================================================
\begin{equation}%                Equation 1
\rho'=\mathcal{E}(\rho).
\label{eq:1}
\end{equation}
%===============================================================================
Mathematically, $\mathcal{E}$ can have \textit{Kraus representation} as
%===============================================================================
\begin{equation}%                Equation 2
\mathcal{E}(\rho)=\sum\nolimits_m {K_m \rho K_m ^ {\dag}  },
\label{eq:2}
\end{equation}
%===============================================================================
where the $K_m$ are \textit{Kraus operators}, with completeness
%===============================================================================
\begin{equation}%                Equation 3
\sum\nolimits_m {K_m ^{\dag}  K_m }  = I.
\label{eq:3}
\end{equation}
%===============================================================================

Quantum operations are important because they can describe both closed and \textit{open} systems, meaning systems of interest that interact with other systems which are either ignored or not fully known.

In the subject of quantum noise, a quantum operation is often referred to as a \textit{quantum channel}.  A good example of a quantum channel is the depolarizing channel,
%===============================================================================
\begin{equation}%                Equation 4
\mathcal{D}(\rho)=p\textstyle{I \over n}+(1-p)\rho,
\label{eq:4}
\end{equation}
%===============================================================================
where $n$ is the dimension of the Hilbert space, $I$ is the identity, and $p$ is a probability $p\in [0,1]$.  Physically, $\mathcal{D}(\rho)$ transmits $\rho$ exactly with probability $1-p$ and replaces it with the maximally mixed state $\textstyle{I \over n}$ with probability $p$.

A Kraus decomposition of $\mathcal{D}(\rho)$ for a single-qubit system ($n=2$) is well-known to be
%===============================================================================
\begin{equation}%                Equation 5
\{ K_m \}  = \{ \sqrt {1 - {\textstyle{{3p} \over 4}}} \sigma _0 ,\sqrt {{\textstyle{p \over 4}}} \sigma _1 ,\sqrt {{\textstyle{p \over 4}}} \sigma _2 ,\sqrt {{\textstyle{p \over 4}}} \sigma _3 \},
\label{eq:5}
\end{equation}
%===============================================================================
where the $\sigma_m$ are Pauli matrices and $\sigma_{0}\equiv I$ \cite{Niel}.  This Kraus decomposition has two special properties; it is both random unitary (RU) and Hilbert-Schmidt (HS) complete, both of which we now briefly review.

A \textit{random unitary} (RU) Kraus decomposition is one for which each of the Kraus operators in the set are proportional to unitary matrices,
%===============================================================================
\begin{equation}%                Equation 6
\text{RU:}\;\; K_m = \sqrt{p_{m}}U_{m} 
\label{eq:6}
\end{equation}
%===============================================================================
where $\sum\nolimits_m {p_m  = 1}$, $p_m  \in [0,1]$, and $U_m^{\dag} = U_{m}^{-1}$.  A quantum channel with an RU decomposition is always correctable since unitary operators are always invertible.

The property of \textit{Hilbert-Schmidt completeness} of a set of operators $\{\eta_{m}\}$ is their ability to expand any operator $A$ in a linear combination of complex coefficients $A_m$ as
%===============================================================================
\begin{equation}%                Equation 7
A = \sum\limits_{m = 0}^{n^2 -1} {A_m \eta _m }.
\label{eq:7}
\end{equation}
%===============================================================================
The $A_m$ are found using the orthogonality of the $\eta_{m}$ under the Hilbert-Schmidt inner product, defined between two operators $A$ and $B$ as $A \cdot B \equiv \text{tr}(A^{\dag} B)$.  Examples of HS-complete operators are the Pauli matrices $\{\sigma_{m}\}$ and the generalized Gell-Mann matrices $\{\lambda_{m}\}$ \cite{Gell,Neem,Gri1}.

Thus, the $K_m$ in (\ref{eq:5}) are both RU and HS complete, two properties which enable the proof that all $n$-level quantum channels are correctable, as we will show in \Sec{III}.

To develop the main results of this paper organically, its presentation will be as follows.  First, in \Sec{II}, we present an $n$-level RU decomposition of all depolarization channels, using both an iterative definition of the full channel, and an explicit method for constructing the RU Kraus operators directly. (The RU-decomposability of depolarization channels is well-known, but the set of errors we derive here is convenient for proving their HS completeness.)  Then, in \Sec{III}, we review the conditions for the correctability of a quantum channel, and show that if any \textit{one} Kraus set has the properties of both correctability and HS-completeness, then that guarantees the correctability of \textit{all} quantum channels.  The fact that the $n$-level RU decomposition of the depolarization channel from \Sec{II} possesses these properties then proves the correctability of all quantum channels.  In \Sec{IV}, we use this discovery to generate a list of requirements for constructing universal error-correction operations.
%                                  END of I
%*******************************************************************************
%*******************************************************************************
%            II. RU Decompositions of All Depolarization Channels
\section{\label{sec:II}RU Decompositions of All Depolarization Channels}
We will use two methods to decompose depolarization channels.  One uses a ``top-down'' approach looking at behaviors of groups of unitary operators on an input state, and the other uses a more direct approach to build the Kraus operators directly.  Since the top-down approach is more intuitive we will look at that method first.
%-------------------------------------------------------------------------------
%                  II.A. Recursive Definition of RU Depolarization
\subsection{\label{sec:II.A}Recursive Definition of RU Depolarization}
To motivate this discussion as naturally as possible, we will break it up into several smaller tasks which will serve to illustrate why the RU decomposition of depolarization channels is always possible with this method.
%...............................................................................
%               II.A.1. $n$-Level RU Maximal Dephasing Channel
\subsubsection{\label{sec:II.A.1}$n$-Level RU Maximal-Dephasing Channel}
First, let the RU maximal-dephasing channel $\Delta(\rho)$ be the channel that sends all off-diagonal elements of $\rho$ to $0$, while preserving its diagonal elements, doing so with RU Kraus operators.  This is given by the recursive formula
%===============================================================================
\begin{equation}%                Equation 8
\begin{array}{*{20}l}
   {\Delta (\rho )} &\!\! { \equiv \Lambda _{n - 1}  \circ  \cdots  \circ \Lambda _1 (\rho )}  \\
   {} &\!\! { \equiv \Lambda _{n - 1} ( \cdots (\Lambda _1 (\rho )) \cdots )},  \\
\end{array}
\label{eq:8}
\end{equation}
%===============================================================================
where $\Lambda_k$ is defined by
%===============================================================================
\begin{equation}%                Equation 9
\Lambda _k (\rho ) \equiv \textstyle{1 \over 2 }\rho  + \textstyle{1 \over 2 }N_k \rho N_k^{\dag},
\label{eq:9}
\end{equation}
%===============================================================================
and $N_k$ is the diagonal unitary matrix
%===============================================================================
\begin{equation}%                Equation 10
N_k  \equiv I - 2E_{(k,k)}^{[n]},
\label{eq:10}
\end{equation}
%===============================================================================
where \smash{$E_{(a,b)}^{[n]} $} is the $n\times n$ \textit{elementary matrix} with a $1$ in its row-$a$, column-$b$ entry and zeros elsewhere.

As an example of how $\Delta(\rho)$ works, we will apply $\Lambda_k (\rho)$ recursively to an arbitrary $4$-level state.  First, note that $N_k$ are just diagonal matrices of ones except that the row-$k$, column-$k$ entry is $-1$, so for example, $N_1  = \text{diag}\{  - 1,1,1,1\}$.  Then, the effect of these on $\rho$ is to flip the sign of the row-$k$, column-$k$ entries, such as
%===============================================================================
\begin{equation}%                Equation 11
N_1 \rho N_1^{\dag}   = \left( {\begin{array}{*{20}r}
   {\rho _{1,1} } & { - \rho _{1,2} } & { - \rho _{1,3} } & { - \rho _{1,4} }  \\
   { - \rho _{2,1} } & {\rho _{2,2} } & {\rho _{2,3} } & {\rho _{2,4} }  \\
   { - \rho _{3,1} } & {\rho _{3,2} } & {\rho _{3,3} } & {\rho _{3,4} }  \\
   { - \rho _{4,1} } & {\rho _{4,2} } & {\rho _{4,3} } & {\rho _{4,4} }  \\
\end{array}} \right)\!.
\label{eq:11}
\end{equation}
%===============================================================================
Thus, adding $\rho$ to this and dividing by $2$ gets rid of all the sign-flipped entires in (\ref{eq:11}).  Performing this operation repeatedly for $k$ up to $n-1$ decoheres the input state as
%===============================================================================
\begin{equation}%                Equation 12
\begin{array}{*{20}l}
   {\Lambda _1 (\rho )} &\!\! {\! =\! \left(\! {\begin{array}{*{20}c}
   {\rho _{1,1} } & 0 & 0 & 0  \\
   0 & {\rho _{2,2} } & {\rho _{2,3} } & {\rho _{2,4} }  \\
   0 & {\rho _{3,2} } & {\rho _{3,3} } & {\rho _{3,4} }  \\
   0 & {\rho _{4,2} } & {\rho _{4,3} } & {\rho _{4,4} }  \\
\end{array}}\! \right)}  \\
   {\Lambda _2 (\Lambda _1 (\rho ))} &\!\! {\! =\! \left(\! {\begin{array}{*{20}c}
   {\rho _{1,1} } & 0 & 0 & 0  \\
   0 & {\rho _{2,2} } & 0 & 0  \\
   0 & 0 & {\rho _{3,3} } & {\rho _{3,4} }  \\
   0 & 0 & {\rho _{4,3} } & {\rho _{4,4} }  \\
\end{array}}\! \right)}  \\
   {\Delta(\rho)\equiv\Lambda _3 (\Lambda _2 (\Lambda _1 (\rho )))} &\!\! {\! =\! \left(\! {\begin{array}{*{20}c}
   {\rho _{1,1} } & 0 & 0 & 0  \\
   0 & {\rho _{2,2} } & 0 & 0  \\
   0 & 0 & {\rho _{3,3} } & 0  \\
   0 & 0 & 0 & {\rho _{4,4} }  \\
\end{array}}\! \right)}\!,  \\
\end{array}
\label{eq:12}
\end{equation}
%===============================================================================
thus removing coherence while preserving probabilities.  This channel was symbolically verified for all dimensions up to $n=13$, and we can claim that it works for all $n$.  
%                                 End of II.A.1
%...............................................................................
%...............................................................................
%          II.A.2. $n$-Level RU Diagonal-Input Randomizer Channel
\subsubsection{\label{sec:II.A.2}$n$-Level RU Diagonal-Input Randomizer Channel}
Now, the next step towards an RU depolarization channel is a method of equalizing all the probabilities of a diagonal state using RU Kraus operators.  This can be accomplished using an RU \textit{permutation channel},
%===============================================================================
\begin{equation}%                Equation 13
\Pi (\rho ) \equiv {\textstyle{1 \over n}}\sum\limits_{m = 1}^n {\Pi _m } \rho \Pi _m^{\dag},
\label{eq:13}
\end{equation}
%===============================================================================
where we use unitary permutation matrices,
%===============================================================================
\begin{equation}%                Equation 14
\Pi _m  \equiv \left( {\begin{array}{*{20}c}
   {R_m }  \\
   {R_{m + 1} }  \\
    \vdots   \\
   {R_{m - 1} }  \\
\end{array}} \right),
\label{eq:14}
\end{equation}
%===============================================================================
where $R_m$ is the $m$th elemental row vector,
%===============================================================================
\begin{equation}%                Equation 15
R_m  \equiv (\begin{array}{*{20}c}
   {0_1 } &  \cdots  & {0_{m - 1} } & {1_m } & {0_{m + 1} } &  \cdots  & {0_n }  \\
\end{array}),
\label{eq:15}
\end{equation}
%===============================================================================
so the rows of $\Pi _m$ start with $R_m$ and cycle through all indices $m$ in increasing order, starting over at $1$ after $n$.  For example, for $n=4$, 
%===============================================================================
\begin{equation}%                Equation 16
\begin{array}{*{20}l}
   {\Pi _1  = \left( {\begin{array}{*{20}c}
   1 & 0 & 0 & 0  \\
   0 & 1 & 0 & 0  \\
   0 & 0 & 1 & 0  \\
   0 & 0 & 0 & 1  \\
\end{array}} \right)\!,} &\!\! {\Pi _2  = \left( {\begin{array}{*{20}c}
   0 & 1 & 0 & 0  \\
   0 & 0 & 1 & 0  \\
   0 & 0 & 0 & 1  \\
   1 & 0 & 0 & 0  \\
\end{array}} \right)\!,}  \\
   {\Pi _3  = \left( {\begin{array}{*{20}c}
   0 & 0 & 1 & 0  \\
   0 & 0 & 0 & 1  \\
   1 & 0 & 0 & 0  \\
   0 & 1 & 0 & 0  \\
\end{array}} \right)\!,} &\!\! {\Pi _4  = \left( {\begin{array}{*{20}c}
   0 & 0 & 0 & 1  \\
   1 & 0 & 0 & 0  \\
   0 & 1 & 0 & 0  \\
   0 & 0 & 1 & 0  \\
\end{array}} \right)\!.}  \\
\end{array}
\label{eq:16}
\end{equation}
%===============================================================================
Thus, any diagonal input to $\Pi (\rho )$ is equalized, such as
%===============================================================================
\begin{equation}%                Equation 17
\Pi (\text{diag}\{ a,b,c,d\} ) = {\textstyle{{a + b + c + d} \over 4}}I.
\label{eq:17}
\end{equation}
%===============================================================================
%                                 End of II.A.2
%...............................................................................
%...............................................................................
%                II.A.3. $n$-Level RU Maximal-Mixing Channel
\subsubsection{\label{sec:II.A.3}$n$-Level RU Maximal-Mixing Channel}
Immediately, we see that if the trace of the diagonal input to $\Pi (\rho )$ is $1$, then $\Pi (\rho )$ will produce the maximally mixed state $\textstyle{I \over n}$.  Then, since $\Delta(\rho)$ always produces a diagonal state for which $\text{tr}(\rho)=1$, we can define an RU maximal-mixing channel as
%===============================================================================
\begin{equation}%                Equation 18
\mathcal{M}(\rho ) \equiv \Pi (\Delta (\rho )) = {\textstyle{I \over n}},
\label{eq:18}
\end{equation}
%===============================================================================
where $\Pi (\rho )$ is given in (\ref{eq:13}), and $\Delta (\rho )$ is given in (\ref{eq:8}).  For example, putting the result of (\ref{eq:12}) into (\ref{eq:18}) produces
%===============================================================================
\begin{equation}%                Equation 19
\mathcal{M}(\Delta(\rho) ) = \left( {\begin{array}{*{20}c}
   {{\textstyle{1 \over 4}}} & 0 & 0 & 0  \\
   0 & {{\textstyle{1 \over 4}}} & 0 & 0  \\
   0 & 0 & {{\textstyle{1 \over 4}}} & 0  \\
   0 & 0 & 0 & {{\textstyle{1 \over 4}}}  \\
\end{array}} \right),
\label{eq:19}
\end{equation}
%===============================================================================
for \textit{all} possible input states $\rho$, whether they are mixed or pure.  Thus, (\ref{eq:18}) produces the maximally mixed state for all $n$-level systems using only RU Kraus operators.
%                                 End of II.A.3
%...............................................................................
%...............................................................................
%                II.A.4. The $n$-Level RU Depolarization Channel
\subsubsection{\label{sec:II.A.4}The $n$-Level RU Depolarization Channel}
Essentially our work is done by (\ref{eq:18}), but to finish the job, note that $\mathcal{D}(\rho)$ in (\ref{eq:4}) involves another term past the maximally mixed state.  However, this feature is trivial because it merely increases the weight of the identity operator term in an RU Kraus decomposition while scaling the others by $p$, which preserves the RU property.

\begin{figure}[H]%                   FIGURE 1
\centering
\includegraphics[width=0.99\linewidth]{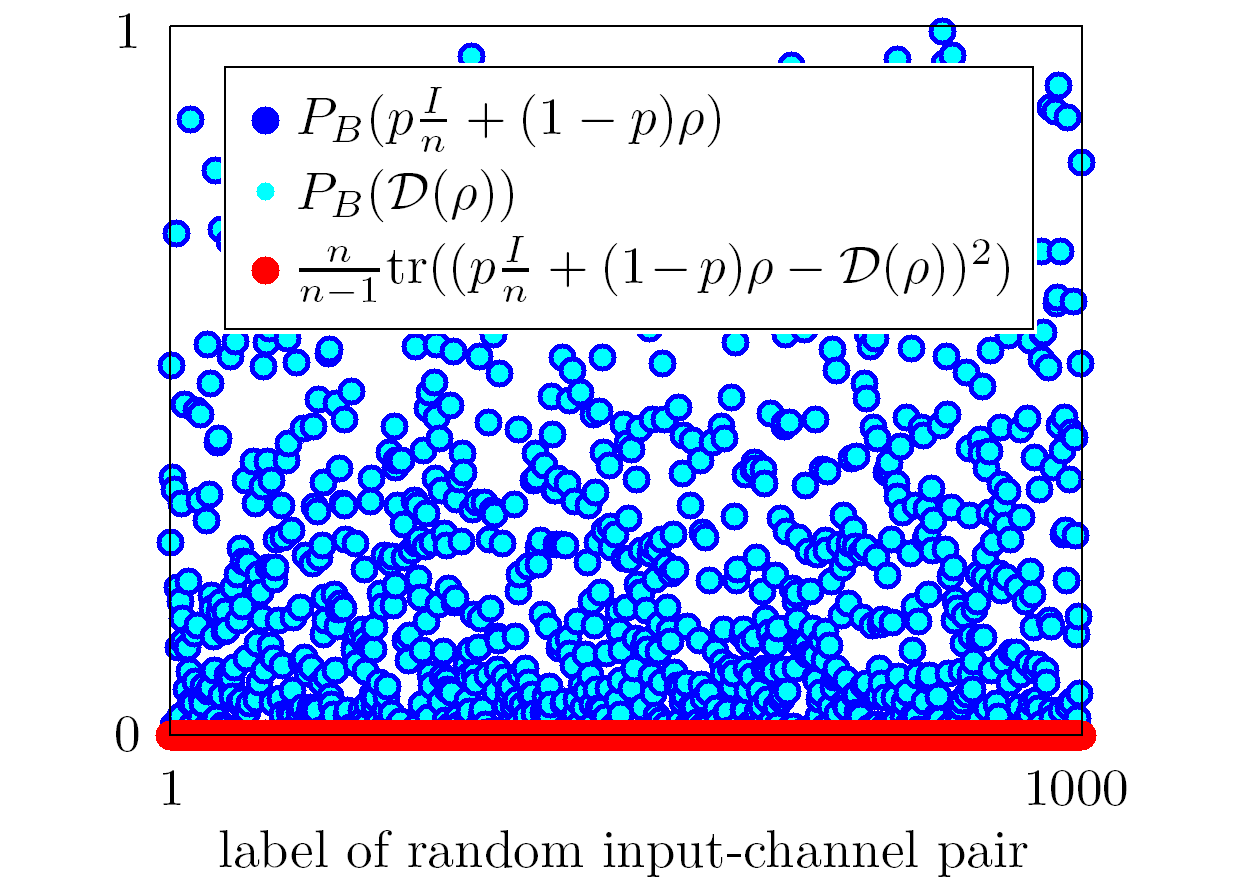}%Base name of figure file here.
\caption[]{\label{fig:1}(color online) Plot of $1000$ random two-qubit depolarization channels $\mathcal{D}(\rho)$ acting on random mixed input states $\rho$ represented two ways each.  First, for visual evidence that different states are used, the height of the dark blue dots is the Bloch purity of the correct output state $P_{B}(p{\textstyle{I \over n}} + (1 - p)\rho)$, where Bloch purity is  $P_{B}(\rho)\equiv (n\text{tr}(\rho^2)-1)/(n-1)$, so that $P_{B}(\rho)=1$ iff $\rho$ is pure, and $P_{B}(\rho)\in [0,1)$ if $\rho$ is mixed, \cite{HedH}.  The height of the light blue dots is the Bloch purity of the RU-reconstructed output state $P_{B}(\mathcal{D}(\rho))$, where $\mathcal{D}(\rho)$ is defined in (\ref{eq:20}).  The \textit{true} necessary and sufficient test is the height of the red dots, which is the square magnitude of the difference of the Bloch vectors of the correct output state and the RU-reconstructed output state $\mathcal{D}(\rho)$, computed as $\frac{n}{n-1}\text{tr}((p{\textstyle{I \over n}} + (1 - p)\rho-\mathcal{D}(\rho))^2)$.  The reconstruction is successful iff the red dot has a height of zero.  Thus, this method works for all states tested.  Note that it is not limited to two qubits; that is merely an arbitrary choice here.}
\end{figure}

Thus, the full RU Kraus decomposition of the $n$-level depolarizing channel is
%===============================================================================
\begin{equation}%                Equation 20
\mathcal{D}(\rho ) = p\mathcal{M}(\rho ) + (1 - p)\rho,
\label{eq:20}
\end{equation}
%===============================================================================
where $\mathcal{M}(\rho )$ is given in (\ref{eq:18}), and the actual RU Kraus operators of $\mathcal{D}(\rho)$ are not immediately visible in (\ref{eq:20}) because of the recursive nature of the functions used to define it.  However, that is acceptable here since the purpose of this section was merely to motivate the procedure.  \Figure{1} demonstrates that the RU depolarization channel $\mathcal{D}(\rho)$ defined in (\ref{eq:20}) truly produces the correct output which is $p{\textstyle{I \over n}} + (1 - p)\rho$ as defined in (\ref{eq:4}).

Next, we will determine the RU Kraus operators directly, which will be more abstract, but more practically useful in many cases.

Also, hereafter we will only focus on the maximal-mixing channel $\mathcal{M}(\rho)$, since we have just shown that if $\mathcal{M}(\rho)$ has an RU decomposition, then $\mathcal{D}(\rho)$ does as well.
%_______________________________________________________________________________
%_______________________________________________________________________________
%                                 End of II.A.4
%...............................................................................
%                                  END of II.A
%-------------------------------------------------------------------------------
%-------------------------------------------------------------------------------
%         II.B. Explicit Kraus-Operator Definition of RU Depolarization
\subsection{\label{sec:II.B}Explicit Kraus-Operator Definition of RU Depolarization}
To develop a direct characterization of the RU Kraus operators of $\mathcal{M}(\rho)$, and by extension the depolarization channel, we will use the iterative procedure above to generate the operators.  We will look at three different methods.  The first is to look at the results of expanding the iterative method directly.  The second is a simple qualitative observation of the first method.  The third is a closed-form method that allows explicit construction of the RU Kraus operators of $\mathcal{M}(\rho)$ without reference to the iterative method.
%...............................................................................
%   II.B.1. Obtaining RU Kraus Operators from Expansion of the Iterative Method
\subsubsection{\label{sec:II.B.1}Obtaining RU Kraus Operators from Expansion of the Iterative Method}
First, due to the definition of the permutation channel $\Pi(\rho)$ in (\ref{eq:13}), we can separate the Kraus operators into permutation groups, based on which permutation operator $\Pi_m$ is being used on the input.

Then, note that $\Delta(\rho)$ from (\ref{eq:8}) produces \textit{twice} the number of Kraus operators needed because half of its Kraus operators are merely different from the others by a global matrix factor of $-1$, which cannot affect the decomposition.  For example, the unitary parts of the Kraus operators of $\Delta(\rho)$ for $n=4$ are
%===============================================================================
\begin{equation}%                Equation 21
\begin{array}{*{20}l}
   {N_0 } &\!\! { \equiv \text{diag}\{ 1,1,1,1\} }  \\
   {N_1 } &\!\! { = \text{diag}\{  - 1,1,1,1\} }  \\
   {N_2 } &\!\! { = \text{diag}\{ 1, - 1,1,1\} }  \\
   {N_3 } &\!\! { = \text{diag}\{ 1,1 - 1,1\} }  \\
   {N_4 } &\!\! { = \text{diag}\{ 1,1,1, - 1\} }  \\
   {N_2 N_1 } &\!\! { = \text{diag}\{  - 1, - 1,1,1\} }  \\
   {N_3 N_1 } &\!\! { = \text{diag}\{  - 1,1, - 1,1\} }  \\
   {N_4 N_1 } &\!\! { = \text{diag}\{  - 1,1,1, - 1\} }  \\
   {N_3 N_2 } &\!\! { = \text{diag}\{ 1, - 1, - 1,1\} }  \\
   {N_4 N_2 } &\!\! { = \text{diag}\{ 1, - 1,1, - 1\} }  \\
   {N_4 N_3 } &\!\! { = \text{diag}\{ 1,1, - 1, - 1\} }  \\
   {N_3 N_2 N_1 } &\!\! { = \text{diag}\{  - 1, - 1, - 1,1\} }  \\
   {N_4 N_2 N_1 } &\!\! { = \text{diag}\{  - 1, - 1,1, - 1\} }  \\
   {N_4 N_3 N_1 } &\!\! { = \text{diag}\{  - 1,1, - 1, - 1\} }  \\
   {N_4 N_3 N_2 } &\!\! { = \text{diag}\{ 1, - 1, - 1, - 1\} }  \\
   {N_4 N_3 N_2 N_1 } &\!\! { = \text{diag}\{  - 1, - 1, - 1, - 1\} },  \\
\end{array}
\label{eq:21}
\end{equation}
%===============================================================================
which include only $2^{n-1}=8$ unique operators.

Then, for the full channel $\mathcal{M}(\rho)$, there are $n$ permutation families of Kraus operators, where the $m$th family is simply $\Pi_m$ left-multiplied to the set of unique $N$-operators, such as the first half of (\ref{eq:21}), resulting in a total of $n2^{n-1}$ unitary operators.  The set is made properly RU by then normalizing with a factor of \smash{$\textstyle{1 \over {\sqrt {n2^{n - 1} } }}$}.  Thus, if we call the half-set of $N$-operators $\{N\}$, then the total set of unique RU Kraus operators for $\mathcal{M}(\rho)$ is
%===============================================================================
\begin{equation}%                Equation 22
\{ M_{k} \}  = \{ {\textstyle{1 \over {\sqrt {n2^{n - 1} } }}}\Pi _1 \{ N\} , \ldots ,{\textstyle{1 \over {\sqrt {n2^{n - 1} } }}}\Pi _n \{ N\} \}.
\label{eq:22}
\end{equation}
%===============================================================================
%                                End of II.B.1
%...............................................................................
%...............................................................................
%     II.B.2. Obtaining RU Kraus Operators from Qualitative Observation
\subsubsection{\label{sec:II.B.2}Obtaining RU Kraus Operators from Qualitative Observation}
A much simpler qualitative way to generate the RU Kraus operators for $\mathcal{M}(\rho)$ is to start with the permutation matrices $\Pi_m$, such as in (\ref{eq:16}), and then simply consider all possible sign distributions among their nonzero elements.  Then, normalizing those by \smash{$\textstyle{1 \over {\sqrt {n2^{n - 1} } }}$} and weeding out the half that are just negatives of the others, one obtains the RU Kraus set for $\mathcal{M}(\rho)$.
%                                End of II.B.2
%...............................................................................
%...............................................................................
%             II.B.3. Explicit Construction of RU Kraus Operators
\subsubsection{\label{sec:II.B.3}Explicit Construction of RU Kraus Operators}
Finally, we have the information we need to explicitly construct the RU Kraus operators of $\mathcal{M}(\rho)$.  First, define the vector-index version of the $N$-operators as
%===============================================================================
\begin{equation}%                Equation 23
N_{\mathbf{x}}  \equiv N_{\mathbf{x}}^{[n]}  \equiv I^{[n]}  - 2\!\sum\limits_{k = 1}^{\text{dim}(\mathbf{x})}\!\! {\text{sgn}(x_k )E_{(x_k ,x_k )}^{[n]} },
\label{eq:23}
\end{equation}
%===============================================================================
where $x_k$ are nonnegative integer components of vector $\mathbf{x}$, and note that $N_0  = I$.  For example, if $\mathbf{x}=(3,2)$, then \smash{$N_{\mathbf{x}}=N_{(3,2)}=I-2E_{(3,3)}-2E_{(2,2)}= N_{3}N_{2}$}.

Next, define the vectorized $n$-choose-$k$ function as
%===============================================================================
\begin{equation}%                Equation 24
\text{nCk}(\mathbf{x},k) \equiv \begin{array}{*{20}l}
   \text{The matrix whose rows are}  \\
   \text{each unique combinations of the}  \\
   \text{elements of \textbf{x} taken $k$ at a time.}  \\
\end{array}
\label{eq:24}
\end{equation}
%===============================================================================
For example, if $n=3$, then
%===============================================================================
\begin{equation}%                Equation 25
\begin{array}{*{20}c}
   {\text{nCk}([1,2,3],1) =\! \left(\! {\begin{array}{*{20}c}
   1  \\
   2  \\
   3  \\
\end{array}}\! \right)\!,\;\;\text{nCk}([1,2,3],2) =\! \left(\! {\begin{array}{*{20}c}
   1 & 2  \\
   1 & 3  \\
   2 & 3  \\
\end{array}}\! \right)\!,}  \\
   {\text{nCk}([1,2,3],3) =\! \left(\! {\begin{array}{*{20}c}
   1 & 2 & 3  \\
\end{array}}\! \right)\!,}  \\
\end{array}
\label{eq:25}
\end{equation}
%===============================================================================
where, notice that by convention, the order of the combinations is taken to be increasing left to right, and counting in the standard format where the right-most digit counts upward to $n$ and then resets as the digit to its left increases by one and so on.

Then, the unique set of $N$-operators is
%===============================================================================
\begin{equation}%                Equation 26
\{ N_\mathbf{x} \} ;\;\;\mathbf{x} \in \{ 0,\{ \text{nCk}([1, \ldots ,n],k)\} _{k = 1}^{k = n} \} |_{\#  = 2^{n - 1} },
\label{eq:26}
\end{equation}
%===============================================================================
where we use set-building notation such that $\{ a,\{ b,c\} \}  \equiv \{ a,b,c\}$, and \smash{$\{ a_k \} _{k = 1}^{k = n}  \equiv \{ a_1 , \ldots, a_n \} $}, and \smash{$\{ a, \ldots \} |_{\#  = z}$} means to stop adding elements to the set after $z$ elements have been included.  Thus, (\ref{eq:26}) instructs us to make all the vectors $\mathbf{x}$ in the set it defines, and then use that set of $2^{n-1}$ vector-indices to populate the set $\{ N_\mathbf{x} \}$.

Finally, a full set of RU Kraus operators for the maximal-mixing channel $\mathcal{M}(\rho)$ is given explicitly by
%===============================================================================
\begin{equation}%                Equation 27
\setlength\fboxsep{4pt}   %Margin around equation.
\setlength\fboxrule{0.5pt}%Width of box lines.
\fbox{$\{ M_{(m,\mathbf{x})} \}  \equiv \{ {\textstyle{1 \over {\sqrt {n2^{n - 1} } }}}\Pi _1 \{ N_\mathbf{x} \} , \ldots ,{\textstyle{1 \over {\sqrt {n2^{n - 1} } }}}\Pi _n \{ N_\mathbf{x} \} \},$}%Actual equation goes in {$$} here.
\label{eq:27}
\end{equation}
%===============================================================================
meaning that we simply left-multiply each member of $\{ N_\mathbf{x} \}$ by each of the permutation matrices $\Pi_m$, and then that group, with each member normalized by \smash{$\textstyle{1 \over {\sqrt {n2^{n - 1} } }}$}, constitutes an RU Kraus decomposition of $\mathcal{M}(\rho)$, and thus can be used to RU-decompose $\mathcal{D}(\rho)$ as well.

For $n=4$, the unitary parts of the RU Kraus operators of $\mathcal{M}(\rho)$ for the first permutation set are
%===============================================================================
\begin{equation}%                Equation 28
\begin{array}{*{20}l}
   {\Pi_1 N_0  = \left(\! {\begin{array}{*{20}r}
   1 & 0 & 0 & 0  \\
   0 & 1 & 0 & 0  \\
   0 & 0 & 1 & 0  \\
   0 & 0 & 0 & 1  \\
\end{array}}\! \right)\!,} &\!\!\! {\Pi_1 N_4  = \left(\! {\begin{array}{*{20}r}
   1 & 0 & 0 & 0  \\
   0 & 1 & 0 & 0  \\
   0 & 0 & 1 & 0  \\
   0 & 0 & 0 & { - 1}  \\
\end{array}}\! \right)}  \\
   {\Pi_1 N_1  = \left(\! {\begin{array}{*{20}r}
   { - 1} & 0 & 0 & 0  \\
   0 & 1 & 0 & 0  \\
   0 & 0 & 1 & 0  \\
   0 & 0 & 0 & 1  \\
\end{array}}\! \right)\!,} &\!\!\! {\Pi_1 N_{(1,2)}  = \left(\! {\begin{array}{*{20}r}
   { - 1} & 0 & 0 & 0  \\
   0 & { - 1} & 0 & 0  \\
   0 & 0 & 1 & 0  \\
   0 & 0 & 0 & 1  \\
\end{array}}\! \right)}  \\
   {\Pi_1 N_2  = \left(\! {\begin{array}{*{20}r}
   1 & 0 & 0 & 0  \\
   0 & { - 1} & 0 & 0  \\
   0 & 0 & 1 & 0  \\
   0 & 0 & 0 & 1  \\
\end{array}}\! \right)\!,} &\!\!\! {\Pi_1 N_{(1,3)}  = \left(\! {\begin{array}{*{20}r}
   { - 1} & 0 & 0 & 0  \\
   0 & 1 & 0 & 0  \\
   0 & 0 & { - 1} & 0  \\
   0 & 0 & 0 & 1  \\
\end{array}}\! \right)}  \\
   {\Pi_1 N_3  = \left(\! {\begin{array}{*{20}r}
   1 & 0 & 0 & 0  \\
   0 & 1 & 0 & 0  \\
   0 & 0 & { - 1} & 0  \\
   0 & 0 & 0 & 1  \\
\end{array}}\! \right)\!,} &\!\!\! {\Pi_1 N_{(1,4)}  = \left(\! {\begin{array}{*{20}r}
   { - 1} & 0 & 0 & 0  \\
   0 & 1 & 0 & 0  \\
   0 & 0 & 1 & 0  \\
   0 & 0 & 0 & { - 1}  \\
\end{array}}\! \right)\!,}  \\
\end{array}
\label{eq:28}
\end{equation}
%===============================================================================
and for the second permutation set,
%===============================================================================
\begin{equation}%                Equation 29
\begin{array}{*{20}l}
   {\Pi_2 N_0  = \left(\! {\begin{array}{*{20}r}
   0 & 1 & 0 & 0  \\
   0 & 0 & 1 & 0  \\
   0 & 0 & 0 & 1  \\
   1 & 0 & 0 & 0  \\
\end{array}}\! \right)\!,} &\!\!\! {\Pi_2 N_4  = \left(\! {\begin{array}{*{20}r}
   0 & 1 & 0 & 0  \\
   0 & 0 & 1 & 0  \\
   0 & 0 & 0 & { - 1}  \\
   1 & 0 & 0 & 0  \\
\end{array}}\! \right)}  \\
   {\Pi_2 N_1  = \left(\! {\begin{array}{*{20}r}
   0 & 1 & 0 & 0  \\
   0 & 0 & 1 & 0  \\
   0 & 0 & 0 & 1  \\
   { - 1} & 0 & 0 & 0  \\
\end{array}}\! \right)\!,} &\!\!\!{\Pi_2 N_{(1,2)}  = \left(\! {\begin{array}{*{20}r}
   0 & { - 1} & 0 & 0  \\
   0 & 0 & 1 & 0  \\
   0 & 0 & 0 & 1  \\
   { - 1} & 0 & 0 & 0  \\
\end{array}}\! \right)}  \\
   {\Pi_2 N_2  = \left(\! {\begin{array}{*{20}r}
   0 & { - 1} & 0 & 0  \\
   0 & 0 & 1 & 0  \\
   0 & 0 & 0 & 1  \\
   1 & 0 & 0 & 0  \\
\end{array}}\! \right)\!,} &\!\!\! {\Pi_2 N_{(1,3)}  = \left(\! {\begin{array}{*{20}r}
   0 & 1 & 0 & 0  \\
   0 & 0 & { - 1} & 0  \\
   0 & 0 & 0 & 1  \\
   { - 1} & 0 & 0 & 0  \\
\end{array}}\! \right)}  \\
   {\Pi_2 N_3  = \left(\! {\begin{array}{*{20}r}
   0 & 1 & 0 & 0  \\
   0 & 0 & { - 1} & 0  \\
   0 & 0 & 0 & 1  \\
   1 & 0 & 0 & 0  \\
\end{array}}\! \right)\!,} &\!\!\! {\Pi_2 N_{(1,4)}  = \left(\! {\begin{array}{*{20}r}
   0 & 1 & 0 & 0  \\
   0 & 0 & 1 & 0  \\
   0 & 0 & 0 & { - 1}  \\
   { - 1} & 0 & 0 & 0  \\
\end{array}}\! \right)\!,}  \\
\end{array}
\label{eq:29}
\end{equation}
%===============================================================================
and for the third permutation set,
%===============================================================================
\begin{equation}%                Equation 30
\begin{array}{*{20}l}
   {\Pi_3 N_0  = \left(\! {\begin{array}{*{20}r}
   0 & 0 & 1 & 0  \\
   0 & 0 & 0 & 1  \\
   1 & 0 & 0 & 0  \\
   0 & 1 & 0 & 0  \\
\end{array}}\! \right)\!,} &\!\!\! {\Pi_3 N_4  = \left(\! {\begin{array}{*{20}r}
   0 & 0 & 1 & 0  \\
   0 & 0 & 0 & { - 1}  \\
   1 & 0 & 0 & 0  \\
   0 & 1 & 0 & 0  \\
\end{array}}\! \right)}  \\
   {\Pi_3 N_1  = \left(\! {\begin{array}{*{20}r}
   0 & 0 & 1 & 0  \\
   0 & 0 & 0 & 1  \\
   { - 1} & 0 & 0 & 0  \\
   0 & 1 & 0 & 0  \\
\end{array}}\! \right)\!,} &\!\!\! {\Pi_3 N_{(1,2)}  = \left(\! {\begin{array}{*{20}r}
   0 & 0 & 1 & 0  \\
   0 & 0 & 0 & 1  \\
   { - 1} & 0 & 0 & 0  \\
   0 & { - 1} & 0 & 0  \\
\end{array}}\! \right)}  \\
   {\Pi_3 N_2  = \left(\! {\begin{array}{*{20}r}
   0 & 0 & 1 & 0  \\
   0 & 0 & 0 & 1  \\
   1 & 0 & 0 & 0  \\
   0 & { - 1} & 0 & 0  \\
\end{array}}\! \right)\!,} &\!\!\! {\Pi_3 N_{(1,3)}  = \left(\! {\begin{array}{*{20}r}
   0 & 0 & { - 1} & 0  \\
   0 & 0 & 0 & 1  \\
   { - 1} & 0 & 0 & 0  \\
   0 & 1 & 0 & 0  \\
\end{array}}\! \right)}  \\
   {\Pi_3 N_3  = \left(\! {\begin{array}{*{20}r}
   0 & 0 & { - 1} & 0  \\
   0 & 0 & 0 & 1  \\
   1 & 0 & 0 & 0  \\
   0 & 1 & 0 & 0  \\
\end{array}}\! \right)\!,} &\!\!\! {\Pi_3 N_{(1,4)}  = \left(\! {\begin{array}{*{20}r}
   0 & 0 & 1 & 0  \\
   0 & 0 & 0 & { - 1}  \\
   { - 1} & 0 & 0 & 0  \\
   0 & 1 & 0 & 0  \\
\end{array}}\! \right)\!,}  \\
\end{array}
\label{eq:30}
\end{equation}
%===============================================================================
and for the fourth permutation set,
%===============================================================================
\begin{equation}%                Equation 31
\begin{array}{*{20}l}
   {\Pi_4 N_0  = \left(\! {\begin{array}{*{20}r}
   0 & 0 & 0 & 1  \\
   1 & 0 & 0 & 0  \\
   0 & 1 & 0 & 0  \\
   0 & 0 & 1 & 0  \\
\end{array}}\! \right)\!,} &\!\!\! {\Pi_4 N_4  = \left(\!{\begin{array}{*{20}r}
   0 & 0 & 0 & { - 1}  \\
   1 & 0 & 0 & 0  \\
   0 & 1 & 0 & 0  \\
   0 & 0 & 1 & 0  \\
\end{array}}\! \right)}  \\
   {\Pi_4 N_1  = \left(\! {\begin{array}{*{20}r}
   0 & 0 & 0 & 1  \\
   { - 1} & 0 & 0 & 0  \\
   0 & 1 & 0 & 0  \\
   0 & 0 & 1 & 0  \\
\end{array}}\! \right)\!,} &\!\!\! {\Pi_4 N_{(1,2)}  = \left(\! {\begin{array}{*{20}r}
   0 & 0 & 0 & 1  \\
   { - 1} & 0 & 0 & 0  \\
   0 & { - 1} & 0 & 0  \\
   0 & 0 & 1 & 0  \\
\end{array}}\! \right)}  \\
   {\Pi_4 N_2  = \left(\! {\begin{array}{*{20}r}
   0 & 0 & 0 & 1  \\
   1 & 0 & 0 & 0  \\
   0 & { - 1} & 0 & 0  \\
   0 & 0 & 1 & 0  \\
\end{array}}\! \right)\!,} &\!\!\! {\Pi_4 N_{(1,3)}  = \left(\! {\begin{array}{*{20}r}
   0 & 0 & 0 & 1  \\
   { - 1} & 0 & 0 & 0  \\
   0 & 1 & 0 & 0  \\
   0 & 0 & { - 1} & 0  \\
\end{array}}\! \right)}  \\
   {\Pi_4 N_3  = \left(\! {\begin{array}{*{20}r}
   0 & 0 & 0 & 1  \\
   1 & 0 & 0 & 0  \\
   0 & 1 & 0 & 0  \\
   0 & 0 & { - 1} & 0  \\
\end{array}}\! \right)\!,} &\!\!\! {\Pi_4 N_{(1,4)}  = \left(\! {\begin{array}{*{20}r}
   0 & 0 & 0 & { - 1}  \\
   { - 1} & 0 & 0 & 0  \\
   0 & 1 & 0 & 0  \\
   0 & 0 & 1 & 0  \\
\end{array}}\! \right)\!,}  \\
\end{array}
\label{eq:31}
\end{equation}
%===============================================================================
all $32$ of which are visibly unitary.  The sum of each of these operations applied to any input $\rho$ as ${\textstyle{1 \over { {n2^{n-1}} }}}\sum_{m,\mathbf{x}}\Pi_{m}N_{\mathbf{x}}\rho(\Pi_{m}N_{\mathbf{x}})^{\dag}$ then constitutes an RU decomposition of $\mathcal{M}(\rho)$, and the output will always be the maximally mixed state $\textstyle{I \over n}$ for all possible input states.
%                                End of II.B.3
%...............................................................................
%...............................................................................
%       II.B.4. Hilbert-Schmidt Completeness of the RU Kraus Operators
\subsubsection{\label{sec:II.B.4}Hilbert-Schmidt Completeness of the RU Kraus Operators}
Since it will be useful to us in the next section, we will show here that this set of RU Kraus operators always contains \textit{at least} enough operators to be HS complete.

First, given the total RU set in (\ref{eq:27}), consider only the first $n$ operators from each permutation set.  Then, expanding each in terms of the elementary matrices, we find that all are related by a single transformation matrix,
%===============================================================================
\begin{equation}%                Equation 32
T \equiv T^{[n]}  \equiv {\textstyle{1 \over {\sqrt {n2^{n - 1} } }}}\left( {\Omega ^{[n]}  - 2\sum\limits_{k = 1}^{n - 1} {E_{(k + 1,k)}^{[n]} } } \right)
\label{eq:32}
\end{equation}
%===============================================================================
where $\Omega ^{[n]} $ is the matrix of all ones, defined as
%===============================================================================
\begin{equation}%                Equation 33
\Omega ^{[n]}  \equiv \sum\limits_{j = 1}^n {\sum\limits_{k = 1}^n {E_{(j,k)}^{[n]} } }.
\label{eq:33}
\end{equation}
%===============================================================================
One finds that each group of elementary matrices that supports the nonzero elements in a given permutation matrix $\Pi_m$ is linearly combined by $T$ to form each of the first $n$ RU Kraus operators of that permutation group.

Now here are the crucial two points.  First, since $T$ has a nonvanishing determinant for all dimensions,
%===============================================================================
\begin{equation}%                Equation 34
\text{det}(T^{[n]} ) = {\textstyle{{2^{n - 1} } \over {(n2^{n - 1} )^n }}} \ne 0\;\;\forall n,
\label{eq:34}
\end{equation}
%===============================================================================
then $T$ is always invertible, allowing us to express the elementary matrices as linear combinations of the RU operators.  Second, since the collection of each of the first $n$ operators of each of the $n$ permutation sets involves all $n^2$ of the unique elementary matrices \smash{$E_{(j,k)}^{[n]}$}, then this plus the invertibility of each group ensures that \textit{all} of the elementary matrices can be expressed using this subset of the RU Kraus operators defined in (\ref{eq:27}).

Since the set of all \smash{$E_{(j,k)}^{[n]}$} are HS complete, and since they can be expanded with a subset of the RU Kraus operators of (\ref{eq:27}), then this means that these RU Kraus operators are HS overcomplete, meaning that they provide more than enough operators to expand any matrix as a linear combination of them.

For example, for $n=4$ the transformation matrix is
%===============================================================================
\begin{equation}%                Equation 35
T = {\textstyle{1 \over {\sqrt {32} }}}\!\left(\!\! {\begin{array}{*{20}r}
   1 & 1 & 1 & 1  \\
   { - 1} & 1 & 1 & 1  \\
   1 & { - 1} & 1 & 1  \\
   1 & 1 & { - 1} & 1  \\
\end{array}} \right)\!,
\label{eq:35}
\end{equation}
%===============================================================================
and the four permutation sets of RU Kraus operators are built from the elementary matrices as
%===============================================================================
\begin{equation}%                Equation 36
\begin{array}{l}
 a\!\left(\! {\begin{array}{*{20}c}
   {\Pi _1 N_0 }  \\
   {\Pi _1 N_1 }  \\
   {\Pi _1 N_2 }  \\
   {\Pi _1 N_3 }  \\
\end{array}}\! \right)\! = T\!\left(\! {\begin{array}{*{20}c}
   {E_{(1,1)}^{[4]} }  \\
   {E_{(2,2)}^{[4]} }  \\
   {E_{(3,3)}^{[4]} }  \\
   {E_{(4,4)}^{[4]} }  \\
\end{array}}\!\! \right)\!,\;a\!\left(\! {\begin{array}{*{20}c}
   {\Pi _2 N_0 }  \\
   {\Pi _2 N_1 }  \\
   {\Pi _2 N_2 }  \\
   {\Pi _2 N_3 }  \\
\end{array}}\! \right)\! = T\!\left(\! {\begin{array}{*{20}c}
   {E_{(4,1)}^{[4]} }  \\
   {E_{(1,2)}^{[4]} }  \\
   {E_{(2,3)}^{[4]} }  \\
   {E_{(3,4)}^{[4]} }  \\
\end{array}}\!\! \right) \\ 
 a\!\left(\! {\begin{array}{*{20}c}
   {\Pi _3 N_0 }  \\
   {\Pi _3 N_1 }  \\
   {\Pi _3 N_2 }  \\
   {\Pi _3 N_3 }  \\
\end{array}}\! \right)\! = T\!\left(\! {\begin{array}{*{20}c}
   {E_{(3,1)}^{[4]} }  \\
   {E_{(4,2)}^{[4]} }  \\
   {E_{(1,3)}^{[4]} }  \\
   {E_{(2,4)}^{[4]} }  \\
\end{array}}\!\! \right)\!,\;a\!\left(\! {\begin{array}{*{20}c}
   {\Pi _4 N_0 }  \\
   {\Pi _4 N_1 }  \\
   {\Pi _4 N_2 }  \\
   {\Pi _4 N_3 }  \\
\end{array}}\!\! \right)\! = T\!\left(\! {\begin{array}{*{20}c}
   {E_{(2,1)}^{[4]} }  \\
   {E_{(3,2)}^{[4]} }  \\
   {E_{(4,3)}^{[4]} }  \\
   {E_{(1,4)}^{[4]} }  \\
\end{array}}\! \right)\!, \\ 
 \end{array}
\label{eq:36}
\end{equation}
%===============================================================================
where \smash{$a=\textstyle{1 \over {\sqrt {32} }}$}.  The invertibility of $T$ and the fact that all $n^2$ elementary matrices appear among the right sides of (\ref{eq:36}) then ensures that we can expand all elementary matrices in terms of our RU Kraus operators, and thus by the HS completeness of the elementary matrices, our RU Kraus operators can be used to expand \textit{any} operator, which proves that this subset of (\ref{eq:27}) is HS complete.

Note that the overcompleteness of the full set of operators from (\ref{eq:27}) is not a problem.  In general, many different combinations of the RU Kraus operators can form a particular \smash{$E_{(j,k)}^{[n]}$}.  For example, suppose we have two different Kraus expansions of a given elementary matrix,
%===============================================================================
\begin{equation}%                Equation 37
E_{(j,k)}^{[n]}  = aM_1  + bM_2 ,\;\;\text{and}\;\;E_{(j,k)}^{[n]}  = cM_3  + dM_4.
\label{eq:37}
\end{equation}
%===============================================================================
Then, given any complex numbers $\alpha,\beta$ such that $\alpha+\beta=1$, we could produce infinite expansions of \smash{$E_{(j,k)}^{[n]}$} as
%===============================================================================
\begin{equation}%                Equation 38
E_{(j,k)}^{[n]}\!  =\! (\alpha\!  +\! \beta )E_{(j,k)}^{[n]}\!  = \alpha (aM_1  + bM_2 ) + \beta (cM_3  + dM_4 ).
\label{eq:38}
\end{equation}
%===============================================================================
Thus, we have a great deal of freedom as to how to involve the RU Kraus operators of (\ref{eq:27}) in the expansion of any given operator.

The importance of all of this will become clearer in the discussion of the correctability of quantum channels.
%                                End of II.B.4
%...............................................................................
%                                 END of II.B
%-------------------------------------------------------------------------------
%                                  END of II
%*******************************************************************************
%*******************************************************************************
%                III. Correctability of All Quantum Channels
\section{\label{sec:III}Correctability of All Quantum Channels}
Now that we have shown that it is always possible to find an HS-complete set of RU Kraus operators for the maximal-mixing and depolarization channels in all dimensions, here we will show that this is sufficient to guarantee the correctability of all quantum channels.

To simplify the argument, we will start by observing several essential facts, and culminate by showing that they prove the above claim.
%-------------------------------------------------------------------------------
%                             III.A. Supporting Facts
\subsection{\label{sec:III.A}Supporting Facts}
%...............................................................................
%   III.A.1. General Conditions for the Correctability of Quantum Error Channels
\subsubsection{\label{sec:III.A.1}General Conditions for the Correctability of Quantum Error Channels}
The essential goal of quantum error correction is to correct for the effects of any given quantum channel on some unknown input state, restoring the system to its original state before the noise channel acted.  Thus, we will refer to this ability of correction or restoration as the \textit{correctability} of a channel or noise.

As in \cite{Niel}, we will not assume any particular error-correction scheme, but rather follow only the bare-minimum of requirements.  Thus, we suppose that if a restoration procedure exists to correct noise from channel $\mathcal{E}$, that it can be represented by a single trace-preserving quantum \textit{recovery} operation $\mathcal{R}$, such that
%===============================================================================
\begin{equation}%                Equation 39
\mathcal{R}(\mathcal{E}(\rho_{C} )) \propto \rho_{C} ,
\label{eq:39}
\end{equation}
%===============================================================================
where \smash{$\rho_{C}\equiv U_{C}\rho U_{C}^{\dag}$} is the encoded version of a generally mixed unknown input state $\rho$, where $U_{C}$ is a unitary encoding operation that encodes the computational basis of $\rho$ into a (possibly) larger Hilbert space, while still preserving the superposition properties of the original $\rho$.  By trace-preserving, we mean that Kraus operators $R_k$ of $\mathcal{R}$ have Kraus completeness \smash{$\sum\nolimits_k {R_k ^{\dag}  R_k }  = I$}.  Note that since the input is unknown, the recovery operation $\mathcal{R}$ should not be dependent on the input state.
%                               End of III.A.1
%...............................................................................
%...............................................................................
%      III.A.2. Existence of a Recovery Operation for a Set of Errors: The Correctability Conditions
\subsubsection{\label{sec:III.A.2}Existence of a Recovery Operation for a Set of Errors: The Correctability Conditions}
From \cite{Niel}, a necessary and sufficient condition for the \textit{existence} of a recovery operation $\mathcal{R}$ that corrects channel $\mathcal{E}$ with Kraus operators $\{ E_k \} $ on quantum code $C$ is that
%===============================================================================
\begin{equation}%                Equation 40
P^{\dag}  E_j^{\dag}  E_k P = \alpha _{j,k} P,
\label{eq:40}
\end{equation}
%===============================================================================
where $\alpha$ is a Hermitian matrix with elements $\alpha _{j,k}$, and \textit{code projector} $P$ is a projector onto the \textit{code space} $C$ such that \smash{$P\rho_{C}P^{\dag}=\rho_C$} are states within $C$, and $P^{\dag}=P$.  We can think of the $\{ E_k \} $ as \textit{errors}, and if $\mathcal{R}$ exists for channel $\mathcal{E}$, then $\{ E_k \} $ is a \textit{correctable set of errors}.
%                               End of III.A.2
%...............................................................................
%...............................................................................
%             III.A.3. Generality of a Recovery Operation
\subsubsection{\label{sec:III.A.3}Generality of a Recovery Operation}
Also from \cite{Niel}, suppose that $\mathcal{R}$ exists and corrects errors $\{ E_k \}$ for channel $\mathcal{E}$.  Then, given any \textit{other} quantum operation $\mathcal{F}$ with errors $\{F_{j}\}$, if each $F_{j}$ is a \textit{linear combination} of the $\{ E_k \}$ such that 
%===============================================================================
\begin{equation}%                Equation 41
F_j  = \sum\nolimits_k {m_{j,k} } E_k,
\label{eq:41}
\end{equation}
%===============================================================================
where the $m_{j,k}$ are complex elements of a generally \textit{nonunitary} matrix $m$, then $\mathcal{R}$ \textit{also corrects the effects of noise $\mathcal{F}$ on code $C$}.
%                               End of III.A.3
%...............................................................................
%...............................................................................
%     III.A.4. Different Kraus Sets of the Same Channel Are Unitarily Related
\subsubsection{\label{sec:III.A.4}Different Kraus Sets of the Same Channel Are Unitarily Related}
Another well-known fact is that given any two different Kraus sets $\{ E_k \}$ and $\{ G_l \}$ that decompose the \textit{same} quantum channel $\mathcal{E}$, if the set with fewer elements is ``padded'' with the necessary number of zero-operators then both sets are unitarily related as
%===============================================================================
\begin{equation}%                Equation 42
E_k  = \sum\nolimits_l {} U_{k,l} G_l ,\;\;\text{or}\;\;G_l  = \sum\nolimits_m {} U_{m,l} ^* E_m ,
\label{eq:42}
\end{equation}
%===============================================================================
where $U$ is a unitary matrix.  Note that Kraus sets from \textit{different} channels are not unitarily related, and in general, one error set may not even have the HS completeness necessary to decompose all or any members of a Kraus set belonging to a different channel.
%                               End of III.A.4
%...............................................................................
%...............................................................................
%    III.A.5. Corollary: Unitarily Related Kraus Sets Share Correctability
\subsubsection{\label{sec:III.A.5}Corollary: Unitarily Related Kraus Sets Share Correctability}
By using the fact that the rows (or columns) of a unitary matrix are orthonormal, one can show that two sets of unitarily related Kraus operators both decompose the same quantum channel.  For example, if $\{ G_l \}$ is unitarily related to $\{ E_k \}$, then
%===============================================================================
\begin{equation}%                Equation 43
\mathcal{E}(\rho ) = \sum\nolimits_k {} E_k \rho E_k ^{\dag }   = \sum\nolimits_l {} G_l \rho G_l ^{\dag }.
\label{eq:43}
\end{equation}
%===============================================================================
Then, if $\{ E_k \}$ is also a correctable set of errors, then channel $\mathcal{E}(\rho )$ is correctable, and thus from (\ref{eq:43}), $\{ G_l \}$  is also a set of correctable errors.  More rigorously, testing the correctability conditions for $\{ G_l \}$ and using (\ref{eq:42}) gives
%===============================================================================
\begin{equation}%                Equation 44
\begin{array}{*{20}l}
   {P^{\dag }  G_j^{\dag }  G_k P} &\!\! { = \sum\nolimits_l {\sum\nolimits_m {U_{l,j} U_{m,k} ^* P^{\dag }  E_l ^{\dag }  E_m P} } }  \\
   {} &\!\! {=\sum\nolimits_l {\sum\nolimits_m {U_{l,j} U_{m,k} ^* \alpha _{l,m} P} } }  \\
   {} &\!\! {=\beta _{j,k} P,}  \\
\end{array}
\label{eq:44}
\end{equation}
%===============================================================================
where \smash{$\beta _{j,k}  \equiv \sum\nolimits_l {\sum\nolimits_m {U_{l,j} U_{m,k} ^* \alpha _{l,m} } }$}, so that $\beta$ is Hermitian, since \smash{$\beta^{\dag}=\beta$} and \smash{$\alpha^{\dag}=\alpha$} from (\ref{eq:40}).  Thus, since $\beta$ is Hermitian, it follows that $\{ G_l \}$ is a correctable set of errors, and this result was guaranteed by both the correctability of $\{ E_k \}$ and its unitary relation to $\{ G_l \}$.
%                               End of III.A.5
%...............................................................................
%                                END of III.A
%-------------------------------------------------------------------------------
%-------------------------------------------------------------------------------
%             III.B. Proof that All Quantum Channels are Correctable
\subsection{\label{sec:III.B}Proof that All Quantum Channels are Correctable}
Now that we have amassed several useful facts, we are ready to put them together to prove the main claim of this paper: that all quantum channels are correctable.
%...............................................................................
%        III.B.1. Proof of the Correctability of All Quantum Channels
\subsubsection{\label{sec:III.B.1}Proof of the Correctability of All Quantum Channels}
The motivation for this proof is actually a corollary to \Sec{III.A.3}.  Simply put, if a correctable set of errors $\{ E_k \}$ is also HS complete, then \textit{any} operator can be expanded as a linear combination of the $E_k$, as in (\ref{eq:41}).  This means that \textit{any other Kraus set $\{ F_j \}$ from any channel $\mathcal{F}$ can be expanded using correctable set $\{ E_k \}$, and therefore $\mathcal{F}$ is also correctable.}  Then, since RU decompositions are always correctable, and since (\ref{eq:27}) proves the existence of RU decompositions that are also HS complete, then we can always find a correctable set of HS-complete Kraus operators for any channel, and therefore any Kraus set is correctable, so all quantum channels are correctable.

Now we shall work this out more rigorously.  Suppose we have a quantum channel
%===============================================================================
\begin{equation}%                Equation 45
\mathcal{F}(\rho ) = \sum\nolimits_j {F_j \rho F_j ^{\dag}  },
\label{eq:45}
\end{equation}
%===============================================================================
with Kraus operators $\{ F_j \}$ for which no recovery operation is known.  Now, suppose that for some \textit{different} channel $\mathcal{E}$, its Kraus operator set $\{ E_k \}$ is both correctable and HS complete.  Then, by their HS completeness, we can expand as $F_j  = \sum\nolimits_k {m_{j,k} E_k }$, so that (\ref{eq:45}) becomes
%===============================================================================
\begin{equation}%                Equation 46
\mathcal{F}(\rho ) = \sum\nolimits_k {} \sum\nolimits_l {H_{k,l} E_k \rho E_l ^{\dag}  } ,
\label{eq:46}
\end{equation}
%===============================================================================
where $H_{k,l}  \equiv \sum\nolimits_j {m_{j,k} m_{j,l} ^* } $ are the elements of a Hermitian matrix $H$ whose Hermiticity is a consequence of the form of the sum that defines it.

Now let $\{ G_r \}$ be a different set of Kraus operators for correctable channel $\mathcal{E}$, so that they are unitarily related to correctable set $\{ E_k \}$, as in (\ref{eq:42}).  Then for some unitary matrix $U$, (\ref{eq:46}) becomes
%===============================================================================
\begin{equation}%                Equation 47
\mathcal{F}(\rho ) = \sum\nolimits_r {} \sum\nolimits_s {} \left( {\sum\nolimits_k {} \sum\nolimits_l {} U_{k,r} H_{k,l} U_{l,s} ^* } \right)G_r \rho G_s ^{\dag}.  
\label{eq:47}
\end{equation}
%===============================================================================
If we set \smash{$U \equiv \epsilon _H ^* $}, where \smash{$\epsilon _H$} is the eigenvector matrix of $H$ so that \smash{$\epsilon _H^{\dag}H\epsilon _H$} is diagonal, then \smash{$U_{k,r}  = (\epsilon _H ^* )_{k,r}  = (\epsilon _H^{\dag}  )_{r,k} $} and \smash{$U_{l,s} ^*  = (\epsilon _H ^* )_{l,s} ^*  = (\epsilon _H )_{l,s} $}, so then
%===============================================================================
\begin{equation}%                Equation 48
 \begin{array}{*{20}l}
   {\sum\nolimits_k {} \sum\nolimits_l {} U_{k,r} H_{k,l} U_{l,s} ^* } &\!\! { = \sum\nolimits_k {} \sum\nolimits_l {} (\epsilon _H^{\dag}  )_{r,k} H_{k,l} (\epsilon _H )_{l,s} }  \\
   {} &\!\! { = (\epsilon _H^{\dag}  H\epsilon _H )_{r,s} }  \\
   {} &\!\! { = \lambda _r \delta _{r,s} ,}  \\
\end{array}
\label{eq:48}
\end{equation}
%===============================================================================
where $\lambda _r$ are the (real) eigenvalues of Hermitian matrix $H$.  Then, putting (\ref{eq:48}) into (\ref{eq:47}) yields
%===============================================================================
\begin{equation}%                Equation 49
\mathcal{F}(\rho ) = \sum\nolimits_r {} \lambda _r G_r \rho G_r ^{\dag}\equiv \sum\nolimits_r {\widetilde{F}_r \rho\widetilde{F}_r^{\dag}} .  
\label{eq:49}
\end{equation}
%===============================================================================
Now, from the fact proved in \Sec{III.A.5} that unitarily-related Kraus sets share correctability, since the Kraus operators \smash{$\widetilde{F}_r  \equiv \sqrt {\lambda _r } G_r$} of (\ref{eq:49}) are proportional to operators of correctable set $\{ G_r \}$, then \smash{$\widetilde{F}_r$} is also a correctable set and therefore we have proven that arbitrary quantum channels $\mathcal{F}(\rho )$ are correctable.

To be more rigorous, applying the correctability conditions to the Kraus operators of (\ref{eq:49}) gives
%===============================================================================
\begin{equation}%                Equation 50
 \begin{array}{*{20}l}
   {P^{\dag}  \widetilde{F}_j^{\dag}  \widetilde{F}_k P} &\!\! { = P^{\dag}  \sqrt {\lambda _j } G_j^{\dag}  \sqrt {\lambda _k } G_k P}  \\
   {} &\!\! { = \sqrt {\lambda _j \lambda _k } \sum\nolimits_s {} \sum\nolimits_r {} (\epsilon _H )_{s,j} ^* (\epsilon _H )_{r,k} P^{\dag}  E_s ^{\dag}  E_r P}  \\
   {} & \!\!{ = \sqrt {\lambda _j \lambda _k } \sum\nolimits_s {} \sum\nolimits_r {} (\epsilon _H )_{s,j} ^* (\epsilon _H )_{r,k} \alpha _{s,r} P}  \\
   {} &\!\! { = \gamma _{j,k} P,}  \\
\end{array}
\label{eq:50}
\end{equation}
%===============================================================================
where $\gamma$ is a Hermitian matrix with elements
%===============================================================================
\begin{equation}%                Equation 51
\gamma _{j,k}  \equiv \sqrt {\lambda _j \lambda _k } \sum\nolimits_s {} \sum\nolimits_r {} (\epsilon _H )_{s,j} ^* (\epsilon _H )_{r,k} \alpha _{s,r},
\label{eq:51}
\end{equation}
%===============================================================================
and again $\alpha$ is Hermitian because $\mathcal{E}$ is correctable.  Since \smash{$\{\widetilde{F}_r\}$} satisfies the correctability conditions, then all quantum channels $\mathcal{F}(\rho )$ are truly correctable.

Thus we have proved the central claim of this paper.  Moreover, applying \Sec{III.A.3} to this result means that any recovery operation $\mathcal{R}$ that corrects the effects of a set $\{E_{k}\}$ that is both correctable and HS complete will \textit{also} correct the effects of \textit{all other quantum operations}.

Then, since (\ref{eq:27}) proves-by-demonstration that there always exists an RU (and therefore correctable) set of HS-complete operators in all dimensions, this guarantees that the recovery $\mathcal{R}$ that corrects the Kraus operators of (\ref{eq:27}) will also correct errors from all other channels.
%                               End of III.B.1
%...............................................................................
%...............................................................................
%  III.B.2. Conversion of Any Kraus Set to a Set of Errors Already Known to be Correctable
\subsubsection{\label{sec:III.B.2}Conversion of Any Kraus Set to a Set of Errors Already Known to be Correctable}
Suppose that $\{ F_j \} $ is some arbitrary set of Kraus operators belonging to channel $\mathcal{F}$ for which a recovery operation is not known.

To find a set of verifiably correctable errors for $\mathcal{F}$, first identify an HS-complete set of correctable Kraus operators $\{ E_k \}$, such as in (\ref{eq:27}).  Then, expand the $F_j$ as
%===============================================================================
\begin{equation}%                Equation 52
F_j  = \sum\nolimits_k {} m_{j,k} E_k,
\label{eq:52}
\end{equation}
%===============================================================================
and then form Hermitian matrix $H$ with elements
%===============================================================================
\begin{equation}%                Equation 53
H_{k,l}  \equiv \sum\nolimits_j {} m_{j,k} m_{j,l} ^*.
\label{eq:53}
\end{equation}
%===============================================================================
Next, define a new set of Kraus operators as
%===============================================================================
\begin{equation}%                Equation 54
G_r  \equiv \sum\nolimits_k {} (\epsilon _H )_{k,r} E_k .
\label{eq:54}
\end{equation}
%===============================================================================
where $\epsilon_{H}$ is the eigenvector matrix of $H$.  Finally, a set of correctable Kraus operators for $\mathcal{F}$ is given by 
%===============================================================================
\begin{equation}%                Equation 55
\{ \widetilde{F}_r \}\equiv \{\sqrt{\lambda_{r}}G_{r}\},
\label{eq:55}
\end{equation}
%===============================================================================
where $\lambda_r$ are the eigenvalues of $H$, the $G_{r}$ are defined in (\ref{eq:54}), and the corresponding known-to-be-correctable decomposition of $\mathcal{F}$ is given in (\ref{eq:49}). 
%                               End of III.B.2
%...............................................................................
%...............................................................................
%           III.B.3. Proof That the Converted Kraus Set is Correctable
\subsubsection{\label{sec:III.B.3}Proof That the Converted Kraus Set is Correctable}
Now, given that (\ref{eq:52}-\ref{eq:55}) shows how to find a correctable set of Kraus operators for \textit{any} channel $\mathcal{F}$, here we will verify that it is actually correctable by finding the recovery operation $\mathcal{R}$ that restores the input state.  This section relies heavily on a less-detailed derivation in \cite{Niel}, but is shown here as a convenient illustration of why the recovery operation works.

Thus, starting with an arbitrary channel $\mathcal{F}$, if its correctable set is $\{ \widetilde{F}_r \}\equiv \{\sqrt{\lambda_{r}}G_{r}\}$ as in (\ref{eq:55}), then the correctability conditions for $\{ \widetilde{F}_r \}$ are $P^{\dag}  \widetilde{F}_k \widetilde{F}_l P = \gamma _{k,l} P$ as given in (\ref{eq:50}), where $\gamma$ is Hermitian, with elements given in (\ref{eq:51}).  Then, to simplify (\ref{eq:50}), define the unitarily related Kraus set,
%===============================================================================
\begin{equation}%                Equation 56
\overline{F}_a  \equiv \sum\nolimits_k {(\epsilon _\gamma ^{\dag}  )_{a,k} ^* \widetilde{F}_k } ,
\label{eq:56}
\end{equation}
%===============================================================================
where $\epsilon _\gamma$ is the eigenvector matrix of $\gamma$ so that $\epsilon _\gamma ^{\dag} \gamma \epsilon _\gamma$ is diagonal.  Then the correctability conditions for $\{\overline{F}_a\}$ are
%===============================================================================
\begin{equation}%                Equation 57
\begin{array}{*{20}l}
   {P^{\dag}  \overline{F}_a ^{\dag}  \overline{F}_b P} &\!\! { = \sum\nolimits_k {\sum\nolimits_l {(\epsilon _\gamma ^{\dag}  )_{a,k} (\epsilon _\gamma ^ {\dag}  )_{b,l} ^* P^{\dag}  \widetilde{F}_k ^{\dag}  \widetilde{F}_l P} } }  \\
   {} &\!\! { = \left( {\sum\nolimits_k {\sum\nolimits_l {(\epsilon _\gamma ^{\dag}  )_{a,k} \gamma _{k,l} (\epsilon _\gamma  )_{l,b} } } } \right)P}  \\
   {} &\!\! { = \left( {\epsilon _\gamma ^{\dag}  \gamma \epsilon _\gamma  } \right)_{a,b} P}  \\
   {} &\!\! { = \delta _{a,b} d_b P,}  \\
\end{array}
\label{eq:57}
\end{equation}
%===============================================================================
where we used (\ref{eq:50}), and $d_b$ are the eigenvalues of $\gamma$.  Next, find the right polar decomposition of $\overline{F}_k P$ as
%===============================================================================
\begin{equation}%                Equation 58
\begin{array}{*{20}l}
   {\overline{F}_k P} &\!\! { = U_k \sqrt {(\overline{F}_k P)^{\dag}  (\overline{F}_k P)} }  \\
   {} &\!\! { = U_k \sqrt {P^{\dag}  \overline{F}_k ^{\dag}  \overline{F}_k P} }  \\
   {} &\!\! { = U_k \sqrt {d_k P} }  \\
   {} &\!\! { = \sqrt {d_k } U_k P}  \\
\end{array}
\label{eq:58}
\end{equation}
%===============================================================================
where we used (\ref{eq:57}) in the second line, and $U_k$ is unitary. Then,  right-multiplying (\ref{eq:58}) by \smash{$U_k^{\dag}$}, we obtain \smash{$\overline{F}_k PU_k ^{\dag}   = \sqrt {d_k } U_k PU_k ^{\dag}   = \sqrt {d_k } P_k $}, with orthogonal projectors \smash{$P_k  \equiv U_k PU_k ^{\dag} $}, from which we find that
%===============================================================================
\begin{equation}%                Equation 59
P_k  = {\textstyle{{\overline{F}_k PU_k ^{\dag}  } \over {\sqrt {d_k } }}},\;\;\text{and}\;\;P_k ^{\dag}   = {\textstyle{{U_k P^{\dag}\overline{F}_k ^{\dag}  } \over {\sqrt {d_k } }}}.
\label{eq:59}
\end{equation}
%===============================================================================

If the syndrome measurement is defined by projectors $P_k$, possibly supplemented by another for completeness so that \smash{$\sum\nolimits_k {P_k }  = I$}, then recovery is accomplished by projective measurement \smash{$P_k^{\dag}$} followed by unitary operation \smash{$U_k ^{\dag}  $}.  Thus, the general recovery channel is
%===============================================================================
\begin{equation}%                Equation 60
\mathcal{R}(\sigma ) = \sum\nolimits_k {} U_k^{\dag}  P_k^{\dag} \sigma P_k U_k,
\label{eq:60}
\end{equation}
%===============================================================================
for input state $\sigma$.  Then, if this recovery operation is for a state $P\rho_{C} P^{\dag}$ in the code space that has experienced channel $\mathcal{F}$ so that the input to the recovery channel is \smash{$\sigma  = \mathcal{F}(P\rho_{C} P^{\dag}) = \sum\nolimits_l {\overline{F}_l P\rho_{C} P^{\dag}\overline{F}_l ^{\dag}  } $}, then the recovery channel acting on this input can be written as
%===============================================================================
\begin{equation}%                Equation 61
\mathcal{R}(\mathcal{F}(P\rho_{C} P^{\dag})) = \sum\nolimits_k {\sum\nolimits_l {U_k^{\dag}  P_k^{\dag} \overline{F}_l P\rho_{C} P^{\dag}\overline{F}_l ^{\dag}  P_k U_k} }.
\label{eq:61}
\end{equation}
%===============================================================================
Then, in the left half of (\ref{eq:61}), using (\ref{eq:59}) and (\ref{eq:57}) produces
%===============================================================================
\begin{equation}%                Equation 62
\begin{array}{*{20}l}
   {U_k^{\dag}  P_k^{\dag} \overline{F}_l P  } &\!\! { = U_k^{\dag}  {\textstyle{{U_k P^{\dag}\overline{F}_k ^{\dag}  } \over {\sqrt {d_k } }}}\overline{F}_l P  }  \\
   {} &\!\! { = U_k^{\dag}  U_k {\textstyle{{P^{\dag}\overline{F}_k ^{\dag}  \overline{F}_l P} \over {\sqrt {d_k } }}}  }  \\
   {} &\!\! { = I{\textstyle{{\delta _{k,l} d_l P} \over {\sqrt {d_k } }}}  }  \\
   {} &\!\! { = \delta _{k,l} \sqrt {d_k } P,  }  \\
\end{array}
\label{eq:62}
\end{equation}
%===============================================================================
and then putting (\ref{eq:62}) into (\ref{eq:61}) reveals that
%===============================================================================
\begin{equation}%                Equation 63
\begin{array}{*{20}l}
   {\mathcal{R}(\mathcal{F}(P\rho_{C} P^{\dag}))} &\!\! { = \sum\nolimits_k {\sum\nolimits_l {(\delta _{k,l} \sqrt {d_k } P) \rho_{C}  (\delta _{k,l} \sqrt {d_k } P  )^{\dag}  } } }  \\
   {} &\!\! { = \sum\nolimits_k {\sum\nolimits_l {\delta _{k,l} \sqrt {d_k } P\rho_{C} P^{\dag}\sqrt {d_k } \delta _{k,l} } } }  \\
   {} &\!\! { = (\sum\nolimits_k {d_k } )P\rho_{C} P^{\dag}}  \\
   {\mathcal{R}(\mathcal{F}(P\rho_{C} P^{\dag}))} &\!\! {\propto P\rho_{C} P^{\dag}, }  \\
\end{array}
\label{eq:63}
\end{equation}
%=============================================================================== 
which shows that the coded input state \smash{$P\rho_{C} P^{\dag}=\rho_{C}$} is recovered by $\mathcal{R}$, despite the effects of noise channel $\mathcal{F}$, where we note that the proportionality constant \smash{$\sum\nolimits_k {d_k }=\text{tr}(\gamma ) $} is eliminated through normalization.  Then, since \smash{$P\rho_{C} P^{\dag}=\rho_{C}$} and \smash{$\rho_{C}\equiv U_{C}\rho U_{C}^{\dag}$}, we recover true input $\rho$ by decoding the recovery output as \smash{$U_{C}^{\dag}\mathcal{R}(\mathcal{F}(P\rho_{C}P^{\dag})) U_{C}=\rho$}. (If an ancilla was used for encoding, we can partial trace over the ancilla at this point.)

Thus we have shown, starting from a \textit{completely arbitrary} channel $\mathcal{F}$ and its action on an arbitrary input state $\rho$ encoded as \smash{$\rho_{C}\equiv U_{C}\rho U_{C}^{\dag}$}, that there exists a recovery operation $\mathcal{R}$ that removes the effects of $\mathcal{F}$ on $\rho_{C}$.  The key step in proving that this is possible for arbitrary channels $\mathcal{F}$ was actually in the last section, in which we outlined how to find a set of correctable Kraus operators for any arbitrary channel.  The existence of the recovery operation $\mathcal{R}$ is then guaranteed as well, as demonstrated explicitly in this section.

It was not necessary for us to use an RU decomposition as in (\ref{eq:27}), but since RU decompositions are always correctable by virtue of the invertibility of unitary operators, it served as a simple means to show that there exist HS-complete sets that are also correctable.

The final and most important step is now to use \Sec{III.A.3} and all subsequent results to develop the requirements for a \textit{single} recovery channel $\mathcal{R}$ that will correct the effects of \textit{all} quantum channels.
%                               End of III.B.3
%...............................................................................
%                                END of III.B
%-------------------------------------------------------------------------------
%                                  END of III
%*******************************************************************************
%*******************************************************************************
%                          IV. APPLICATIONS
\section{\label{sec:IV}Applications}
The ultimate intention of this paper is in fact the application of the ideas presented here to accomplish recovery from any quantum channel acting on a discrete system.  Since the proof given in the previous section implies that it is \textit{always} possible to recover from the effects of any quantum channel, then the most general application is the construction of \textit{universal recovery channels} $\mathcal{R}$.  Such channels are not unique and depend both upon the quantum code used and the HS-complete correctable Kraus set chosen to expand the arbitrary errors.  Here we present conditions for constructing universal recovery channels, as well as an example of how to verify that any set of errors is correctable.  Finally we look at a special case showing that technically all channels have a state-dependent RU decomposition.  Though not useful for error correction, this fact mathematically agrees with the the result we just proved for state-independent decompositions, that all channels are correctable.
%-------------------------------------------------------------------------------
%        IV.A. Conditions for Constructing Universal Recovery Operations
\subsection{\label{sec:IV.A}Conditions for Constructing Universal Recovery Operations}
Given any $n$-level state \smash{$\rho^{(S)}$} of system $S$ subject to any quantum noise, a universal procedure to protect and recover it after the noise is as follows.  First, prepare an ancilla $A$ in a convenient $n_A$-level pure state \smash{$|\psi_0^{(A)}\rangle$} in product with \smash{$\rho^{(S)}$} as \smash{$\rho^{(S)}\otimes|\psi_0^{(A)}\rangle\langle\psi_0^{(A)}|$}, where \smash{$n_A \geq n^2$} since the minimum number of operators to have HS completeness for an $n$-level system is $n^2$.  Then, find a set of \smash{$n_E \geq n^2$} Kraus operators $\{Q_{m}\}$, with Kraus completeness \smash{$\sum\nolimits_{m}Q_m^{\dag}Q_{m}=I^{[n_{Q}]}$}, where \smash{$n\leq n_Q \leq nn_A$}, that satisfy the following three conditions:
\begin{itemize}
\item[$1.$]$\{ Q_m \}$ must be \textit{Hilbert-Schmidt (HS) complete}, meaning that all matrices $C$ of \smash{$n_Q$} levels can be expanded as \smash{$C = \sum\nolimits_m {c_m Q_m }$} where the scalar coefficients $c_m$ are determined by the orthogonality of $\{ Q_m \}$ under the HS inner product $A \cdot B \equiv \text{tr}(A^{\dag} B)$.
\item[$2.$]$\{ Q_k \}$ must be \textit{correctable}, which means that \smash{$P^{\dag}Q_l ^{\dag}  Q_m P = \alpha _{l,m} P$  $\forall l,m$} such that \smash{$\alpha^{\dag}=\alpha$}, for some joint-system rank-$n$-or-greater projector $P$ such that $P^2  = P$, $\text{tr}(P) \ne 0$, and chosen such that $\alpha^{\dag}=\alpha$ and \smash{$\epsilon _\alpha ^{\dag} = \epsilon _\alpha^{-1}$}, where $\epsilon_\alpha$ is the eigenvector matrix of $\alpha$ such that $\epsilon_{\alpha}^{\dag}\alpha\epsilon_\alpha$ is diagonal.
\item[$3.$]A joint-system unitary encoding operator $U_C$ must be chosen such that the encoded state \smash{$\rho_{C}\equiv U_{C}(\rho^{(S)}\!\otimes|\psi_0^{(A)}\rangle\!\langle\psi_0^{(A)}|)U_{C}^{\dag}$} is unaffected by code projector $P$, so that \smash{$P\rho_{C}P^{\dag}=\rho_{C}$}.  In particular, $U_C$ and $P$ need to be chosen such that they preserve the superposition properties of \smash{$\rho^{(S)}$}.
\end{itemize}

A candidate $P$ can be tested for compatibility with a given $\{Q_m \}$ by building the matrix $\alpha$ with elements
%===============================================================================
\begin{equation}%                  Equation 64
\alpha _{l,m}  \equiv {\textstyle{1 \over {\text{tr}(P)}}}\text{tr}(P^{\dag}Q_l ^{\dag}  Q_m P),
\label{eq:64}
\end{equation}
%===============================================================================
and checking that $\alpha^{\dag}=\alpha$ and $\epsilon _\alpha ^{\dag} \epsilon _\alpha  = I $.  The general procedure for correcting arbitrary errors on an $n$-level system is given by the recovery channel,
%===============================================================================
\begin{equation}%                  Equation 65
\mathcal{R}(\sigma ) \equiv \sum\limits_{k = 1}^{r_{\alpha} + 1} R_k \sigma R_k ^{\dag}  
\label{eq:65}
\end{equation}
%===============================================================================
where $r_{\alpha}\equiv\text{rank}(\alpha)$, and Kraus operators $R_k$ belong to the joint system $(S,A)$ of $nn_A$ levels formed by the $n$-level system of interest $(S)$ and an ancilla $(A)$ of $n_{A}\geq n^2$ levels, and $\sigma\equiv\mathcal{E}(\rho_{C})$ is any error-corrupted state in the code space.  The recovery operators can be written as
%===============================================================================
\begin{equation}%                  Equation 66
R_k  \equiv U_k^{\dag} P_k^{\dag},
\label{eq:66}
\end{equation}
%===============================================================================
with orthogonal projection operators 
%===============================================================================
\begin{equation}%                  Equation 67
P_k  \equiv \left\{ {\begin{array}{*{20}l}
   {{\textstyle{1 \over {\sqrt {\eta _k } }}}\overline{Q}_k PU_k^ {\dag} ;} &\;\; {k \in 1, \ldots ,r_{\alpha} }  \\
   {I^{(S,A)}  - \sum\nolimits_{m = 1}^{r_{\alpha} } {P_m }; } &\;\; {k = r_{\alpha}  + 1,}  \\
\end{array}} \right.
\label{eq:67}
\end{equation}
%===============================================================================
such that $P_k^{\dag}=P_{k}$ and $P_{k}P_{l}=\delta_{k,l}P_{k}$, and where
%===============================================================================
\begin{equation}%                  Equation 68
U_{k \in 1, \ldots ,r_{\alpha} }  \equiv W_k X_k^{\dag}\;\;\;\text{and}\;\;\;U_{r_{\alpha}  + 1},  \equiv I^{(S,A)}
\label{eq:68}
\end{equation}
%===============================================================================
where \smash{$W_k$} and \smash{$X_k$} are unitaries from the singular value decomposition of \smash{$\overline{Q}_k P$} such that \smash{$\overline{Q}_k P = W_k \Sigma_k X_k^{\dag}$} where \smash{$\Sigma_k$} is the matrix of singular values.  The \smash{$\eta_k$} are descending-order eigenvalues \smash{$\eta_1 \geq \cdots \geq\eta_{r_{\alpha}}$} of the matrix $\alpha$.  The \smash{$\overline{Q}_k$} are given by
%===============================================================================
\begin{equation}%                  Equation 69
\overline{Q}_k  \equiv \sum\nolimits_{m = 1}^{n_E } {(\epsilon _\alpha ^{\dag}  )^{*}_{k,m} Q_m } 
\label{eq:69}
\end{equation}
%===============================================================================
where again, $\epsilon_\alpha$ is the eigenvector matrix of $\alpha$ such that $\epsilon_{\alpha}^{\dag}\alpha\epsilon_\alpha$ is diagonal.  The initial state of $S$ is fully recovered by decoding and tracing over the ancilla as
%===============================================================================
\begin{equation}%                  Equation 70
\rho ^{(S)} = \text{tr}_A (U_C^{\dag}  \mathcal{R}(\mathcal{E}(\rho_{C}))U_C ).
\label{eq:70}
\end{equation}
%===============================================================================

Thus, the creation of universal recovery operations has been reduced to judicious choice of $P$ and $U_C$ such that these conditions are met.  The main innovation here is the requirement of the use of an HS-complete set \smash{$\{Q_{m}\}$}.  If \smash{$\{P_{k\in 1\ldots,r_{\alpha}}\}$} is complete \smash{$\sum\nolimits_{k}P_{k}=I^{(S,A)}$}, then the sum's upper limit in (\ref{eq:65}) is $r_\alpha$, and the $(r_{\alpha}+1)$-case in (\ref{eq:67}) and (\ref{eq:68}) is not needed.  Ancillas may not be needed, and it is currently unknown whether global joint-system errors are correctable, but this method \textit{can} be used to construct $n$-level local correction schemes.  Although we give no examples of explicit universal recovery operations here, it is hoped that these steps can help guide their creation.
%                                  END of IV.A
%-------------------------------------------------------------------------------
%-------------------------------------------------------------------------------
%  IV.B. Example of How to Verify the Correctability of an Arbitrary Set of Errors
\subsection{\label{sec:IV.B}Example of How to Verify the Correctability of an Arbitrary Set of Errors}
Here we illustrate the steps in (\ref{eq:52}-\ref{eq:55}) for a specific channel and a set of Kraus operators that we do not know are correctable simply by looking at them.  This actually works for \textit{any} set of Kraus operators for \textit{any} channel, but we use a specific example here as a demonstration.

Consider the channel $\mathcal{F}$ with non-RU Kraus operators
%===============================================================================
\begin{equation}%                Equation 71
\begin{array}{*{20}l}
   {F_1 } &\!\! { = \text{diag}\{ p^2 ,p,p,1\} }  \\
   {F_2 } &\!\! { = \text{diag}\{ pq,0,q,0\} }  \\
   {F_3 } &\!\! { = \text{diag}\{ qp,q,0,0\} }  \\
   {F_4 } &\!\! { = \text{diag}\{ q^2 ,0,0,0\}, }  \\
\end{array}
\label{eq:71}
\end{equation}
%===============================================================================
where $q \equiv \sqrt {1 - p^2 }$, and $p\in[0,1]$ is time-dependent, and this channel models Ornstein-Ulenbeck phase noise, the details of which can be found in \cite{Yu15}, and the output of the channel acting on an arbitrary initial state $\rho(0)$ is 
%===============================================================================
\begin{equation}%                Equation 72
\mathcal{F}(\rho(0)) =\! \left(\! {\begin{array}{*{20}l}
   {\rho _{1,1} (0)} & {\rho _{1,2} (0)p} & {\rho _{1,3} (0)p} & {\rho _{1,4} (0)p^2 }  \\
   {\rho _{2,1} (0)p} & {\rho _{2,2} (0)} & {\rho _{2,3} (0)p^2 } & {\rho _{2,4} (0)p}  \\
   {\rho _{3,1} (0)p} & {\rho _{3,2} (0)p^2 } & {\rho _{3,3} (0)} & {\rho _{3,4} (0)p}  \\
   {\rho _{4,1} (0)p^2 } & {\rho _{4,2} (0)p} & {\rho _{4,3} (0)p} & {\rho _{4,4} (0)}  \\
\end{array}}\! \right)\!,
\label{eq:72}
\end{equation}
%===============================================================================
showing that this a kind of phase damping.  

Observation of (\ref{eq:71}) shows that the $F_j$ are not RU.  Beyond that, to see if this set of errors is correctable, we might try testing the correctablity conditions of (\ref{eq:40}).  However, for that we need some way to choose the code projector $P$, otherwise that test is not very useful.

This is where the steps of (\ref{eq:52}-\ref{eq:55}) become useful.  They allow us to find a \textit{different} set of Kraus operators \smash{$\{\widetilde{F}_{r}\}$} for the \textit{same} channel $\mathcal{F}$, where the new set has the desired property of being able to satisfy the correctability conditions of (\ref{eq:40}), thus guaranteeing that $\mathcal{F}$ is correctable.

First, applying (\ref{eq:52}), if we define the known-correctable errors as the RU set from (\ref{eq:27}), then we find we only need five of them, which we label as \smash{$E_1 \equiv \textstyle{1 \over {\sqrt {32} }}\Pi_{1}N_{0}$}, \smash{$E_2 \equiv \textstyle{1 \over {\sqrt {32} }}\Pi_{1}N_{1}$}, \smash{$E_3 \equiv \textstyle{1 \over {\sqrt {32} }}\Pi_{1}N_{2}$}, \smash{$E_4 \equiv \textstyle{1 \over {\sqrt {32} }}\Pi_{1}N_{3}$}, and \smash{$E_5 \equiv \textstyle{1 \over {\sqrt {32} }}\Pi_{1}N_{4}$}.  Then the given \smash{$\{F_{j}\}$} can be written as
%===============================================================================
\begin{equation}%                Equation 73
\begin{array}{*{20}l}
   {F_1 } &\!\! { = {\textstyle{a \over 2}}(p + 1)^2 E_1  - {\textstyle{a \over 2}}p^2 E_2  - {\textstyle{a \over 2}}pE_3  - {\textstyle{a \over 2}}pE_4  - {\textstyle{a \over 2}}E_5 }  \\
   {F_2 } &\!\! { = {\textstyle{a \over 2}}(p + 1)qE_1  - {\textstyle{a \over 2}}pqE_2  + 0E_3  - {\textstyle{a \over 2}}qE_4  + 0E_5 }  \\
   {F_3 } &\!\! { = {\textstyle{a \over 2}}(p + 1)qE_1  - {\textstyle{a \over 2}}pqE_2  - {\textstyle{a \over 2}}qE_3  + 0E_4  + 0E_5 }  \\
   {F_4 } &\!\! { = {\textstyle{a \over 2}}q^2 E_1  - {\textstyle{a \over 2}}q^2 E_2  + 0E_3  + 0E_4  + 0E_5 ,}  \\
\end{array}
\label{eq:73}
\end{equation}
%===============================================================================
where $a\equiv\sqrt{32}$, which produces the coefficient matrix,
%===============================================================================
\begin{equation}%                Equation 74
m = {\textstyle{a \over 2}}\!\left( {\begin{array}{*{20}c}
   {(p + 1)^2 } & { - p^2 } & { - p} & { - p} & { -1 }  \\
   {(p + 1)q} & { - pq} & 0 & { - q} & 0  \\
   {(p + 1)q} & { - pq} & { - q} & 0 & 0  \\
   {q^2 } & { - q^2 } & 0 & 0 & 0  \\
\end{array}} \right)\!.
\label{eq:74}
\end{equation}
%===============================================================================
Then, setting $s \equiv 1 + p$, putting (\ref{eq:74}) into (\ref{eq:53}), and eliminating $q$ by its definition, we get the Hermitian matrix,
%===============================================================================
\begin{equation}%                Equation 75
H = 8\!\left( {\begin{array}{*{20}c}
   {4s^2 } & { - s^2 } & { - s^2 } & { - s^2 } & { - s^2 }  \\
   { - s^2 } & 1 & p & p & {p^2 }  \\
   { - s^2 } & p & 1 & {p^2 } & p  \\
   { - s^2 } & p & {p^2 } & 1 & p  \\
   { - s^2 } & {p^2 } & p & p & 1  \\
\end{array}} \right)\!.
\label{eq:75}
\end{equation}
%===============================================================================

\begin{figure}[H]%                   FIGURE 2
\centering
\includegraphics[width=0.99\linewidth]{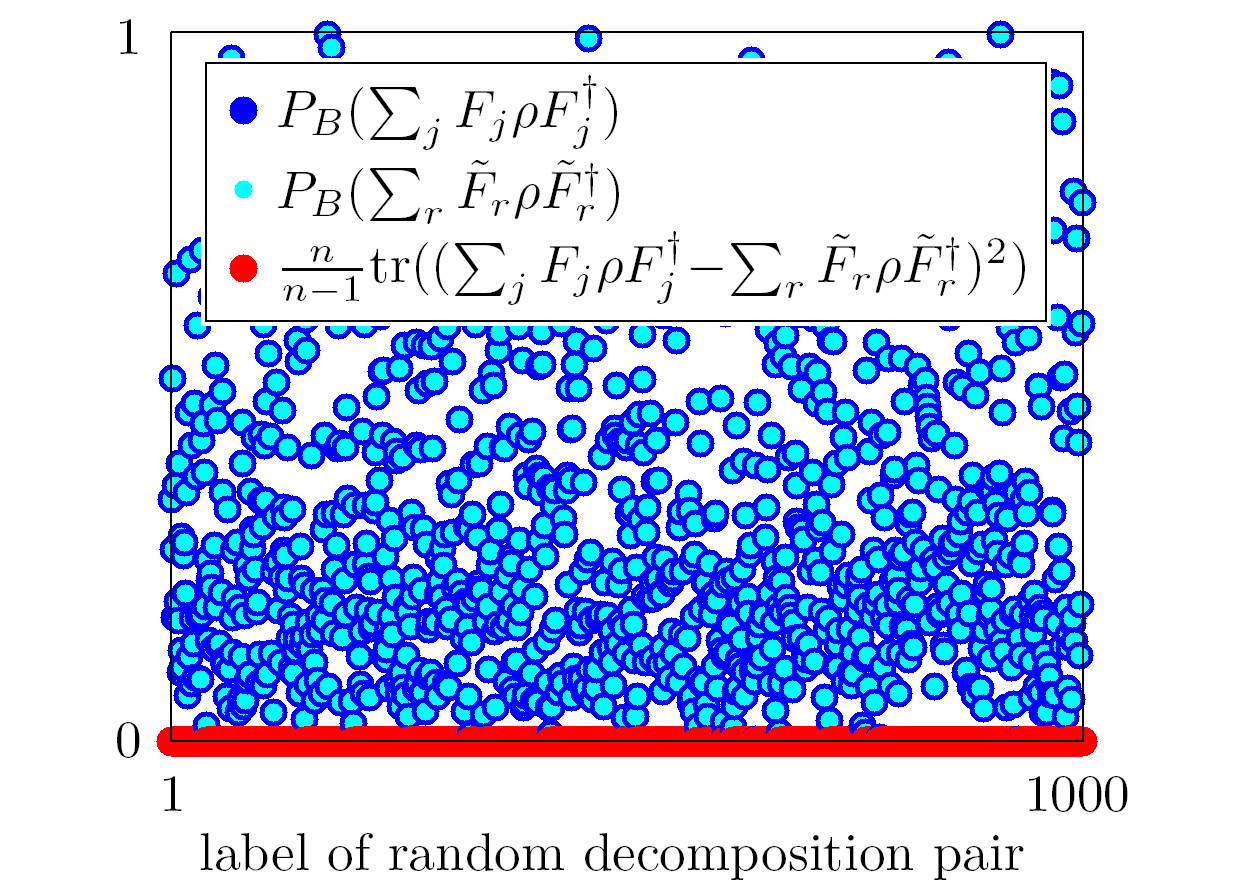}%Base name of figure file here.
\caption[]{\label{fig:2}(color online) Plot of the two-qubit channel $\mathcal{F}$ computed two ways for $1000$ different values of $p\in[0,1]$.  First, for visual evidence that different mixed input states $\rho\equiv\rho(0)$ and probabilities $p$ are used, the dark blue dots are the Bloch purities $P_{B}$ (defined in \Fig{1}) of the output of $\mathcal{F}$ using the $\{F_{j}\}$ of (\ref{eq:71}).  The light blue dots are the Bloch purities of the quantum operation using \smash{$\{\widetilde{F}_{r}\}$} obtained by putting (\ref{eq:75}) into (\ref{eq:54}) and (\ref{eq:55}). However, the \textit{true} necessary and sufficient test is the red dots, which are the square magnitudes of the difference of the Bloch vectors of two differently-computed output states, computed as \smash{$\frac{n}{n-1}\text{tr}((\sum_{j}F_{j}\rho F_{j}^{\dag}-\sum_{r}\widetilde{F}_{r}\rho \widetilde{F}_{r}^{\dag})^2)$}.  The two decompositions of $\mathcal{F}$ produce the same output iff the red dot has a height of zero.  Since all pairs of decompositions produce the exact same output, this gives us good confidence that \smash{$\{\widetilde{F}_{r}\}$} is a valid Kraus decomposition of $\mathcal{F}$.}
\end{figure}

The new set of Kraus operators \smash{$\{\widetilde{F}_{r}\}$} is then obtained from (\ref{eq:54}) and (\ref{eq:55}) by using the eigenvector matrix $\epsilon_{H}$ of $H$ and the eigenvalues $\lambda_r$ of $H$.  At this point, we must abandon symbolic representation and test numerical values.  Therefore, by choosing random values for $p\in [0,1]$, and choosing arbitrary input states $\rho(0)$, we can test that the output states are the same, for both \smash{$\{F_{j}\}$} and \smash{$\{\widetilde{F}_{r}\}$}, which is tested by \Fig{2}.

Now that \Fig{2} has given us confidence that the new Kraus set \smash{$\{\widetilde{F}_{r}\}$} is a valid decomposition of $\mathcal{F}$, it is easy to prove that $\mathcal{F}$ is correctable.  The correctability conditions of \smash{$\{\widetilde{F}_{r}\}$} are given in (\ref{eq:50}), given a known-correctable set $\{E_{k}\}$, as \smash{$P^{\dag}\widetilde{F}_{j}^{\dag}\widetilde{F}_{k}P=\gamma_{j,k}P$}, where the necessary and sufficient condition for correctability is satisfied because $\gamma$ is Hermitian, since its elements are
%===============================================================================
\begin{equation}%                Equation 76
\gamma _{j,k}  = \sqrt {\lambda _j \lambda _k } \sum\nolimits_s {\sum\nolimits_r {(\epsilon _H )_{s,j}^* (\epsilon _H )_{r,k} \alpha _{s,r} } },
\label{eq:76}
\end{equation}
%===============================================================================
where to derive this, we assumed that we already knew the correctability conditions for the known-correctable set $\{E_{k}\}$.  In this case, since $\{E_{k}\}$ is a subset of the HS-complete RU set from (\ref{eq:27}), we \text{know} that it is correctable because every operator is proportional to a unitary operator, so $\alpha$ is Hermitian and so $\gamma$ is also.  Therefore the \smash{$\{\widetilde{F}_{r}\}$} are correctable and thus $\mathcal{F}$ is correctable.
%_______________________________________________________________________________
%_______________________________________________________________________________

To summarize what we just did, we started with a set of Kraus operators that we were not sure was correctable just by looking at them.  Then, we used the steps of (\ref{eq:52}-\ref{eq:55}) to find a different set of Kraus operators for the same channel.  This new set is unitarily related to the set of HS-complete RU Kraus operators from (\ref{eq:27}), and therefore, they inherit the correctability of that set, according to \Sec{III.A.5}.  The fact that we used a specific example here is just for illustration purposes.  This proof is possible for \textit{any} quantum channel, as we saw in \Sec{III.B}.
%                                  END of IV.B
%-------------------------------------------------------------------------------
%-------------------------------------------------------------------------------
%     IV.C. Special Topic: State-Dependent RU Decomposition for All Channels
\subsection{\label{sec:IV.C}Special Topic: State-Dependent RU Decomposition for All Channels}
Here, we address a related, but different topic.  Note that this section is not necessary for anything we have discussed so far.  However, its relevance arises from the fact that it proves that \textit{all} quantum channels can be expressed as RU Kraus expansions, and are therefore correctable, in principle.  The catch is that such RU expansions are generally state-dependent in terms of \textit{both} the input state and the output state of a given quantum channel.

Despite the state-dependence of this method, the fact that all possible mixed output states can be expressed as RU decompositions of all possible pure input states is incredibly powerful in that it suggests, at least theoretically, that any quantum channel's effects are correctable since we can \textit{always} obtain an RU decomposition.  

However, the state-dependence of this decomposition prevents it from being useful for general correction schemes in which we do not know the input state.  Nevertheless, we will propose an application for it beyond its theoretical value.  Now, we consider this universal RU decomposition method.

First, note that any quantum channel $\mathcal{E}$ is just a mapping of some input state $\rho$ to an output state $\rho'$, as
%===============================================================================
\begin{equation}%                Equation 77
\mathcal{E}(\rho ) = \rho '.
\label{eq:77}
\end{equation}
%===============================================================================
Since both states are physical, both have spectral decompositions, and thus, given a particular input-output pair $\{\rho,\rho'\}$ of $n$-level states, we can define the \textit{spectral decomposition of the channel} $\mathcal{E}$ as being that of $\rho'$,
%===============================================================================
\begin{equation}%                Equation 78
\mathcal{E}(\rho ) = \rho ' = \sum\limits_{j = 1}^R {\lambda _j \rho _j '} ,
\label{eq:78}
\end{equation}
%===============================================================================
where \smash{$\rho _j ' \equiv |e_j \rangle \langle e_j |$} are the pure eigenstates of $\rho'$, with corresponding eigenvalues $\lambda _j$ such that \smash{$\sum\nolimits_{j = 1}^n {\lambda _j  = 1} $} and $\lambda _j  \in [0,1]$, and $R \equiv \text{rank}(\rho ')$, where we use the descending order convention for the eigenvalues, $\lambda_{1}\ge\cdots\ge\lambda_{n}$.

Then, given pure state input $\rho \equiv |\psi \rangle \langle \psi |$, since all pure states have the same diagonal matrix of eigenvalues $D$, we can eliminate this using
%===============================================================================
\begin{equation}%                Equation 79
\epsilon _{\rho _j '} ^{\dag}  \rho _j '\epsilon _{\rho _j '}  = D = \epsilon _\rho  ^{\dag}  \rho \epsilon _\rho  ,
\label{eq:79}
\end{equation}
%===============================================================================
where $\epsilon_{A}$ is the unitary eigenvector matrix of $A$, such that \smash{$\epsilon_{A}^{\dag}A\epsilon_{A}$} is diagonal.  Then, solving (\ref{eq:79}) for the channel eigenstates \smash{$\rho _j '$} yields
%===============================================================================
\begin{equation}%                Equation 80
\rho _j ' = \epsilon _{\rho _j '} \epsilon _\rho  ^{\dag}  \rho \epsilon _\rho  \epsilon _{\rho _j '} ^{\dag}   = U_j \rho U_j ^{\dag} ,
\label{eq:80}
\end{equation}
%===============================================================================
where the unitary matrix $U_j$ that converts $\rho$ to $\rho_{j}'$ is
%===============================================================================
\begin{equation}%                Equation 81
U_j  \equiv \epsilon _{\rho _j '} \epsilon _\rho  ^{\dag} .
\label{eq:81}
\end{equation}
%===============================================================================
Thus, putting (\ref{eq:80}) into (\ref{eq:78}) yields the RU Kraus decomposition of $\mathcal{E}(\rho )$ as
%===============================================================================
\begin{equation}%                Equation 82
\mathcal{E}(\rho ) = \sum\limits_{j = 1}^R {\lambda _j U_j \rho U_j ^{\dag}  }  = \sum\limits_{j = 1}^R {} K_j \rho K_j ^{\dag},
\label{eq:82}
\end{equation}
%===============================================================================
where the RU Kraus operators are
%===============================================================================
\begin{equation}%                Equation 83
K_j  \equiv \sqrt {\lambda _j } U_j  = \sqrt {\lambda _j } \epsilon _{\rho _j '} \epsilon _\rho  ^{\dag}.  
\label{eq:83}
\end{equation}
%===============================================================================

Thus, we have proven that given a known pure state input $\rho \equiv |\psi \rangle \langle \psi |$, and a quantum channel $\mathcal{E}(\rho )$ producing state $\rho'$ from input $\rho$, it is \textit{always} possible to obtain an RU Kraus decomposition of $\mathcal{E}(\rho )$.

Physically, this means that the action of any quantum channel on a pure-state input can be thought of as applying each of a set of $R$ unitary operations $U_j$ on input $\rho \equiv |\psi \rangle \langle \psi |$ with probability $\lambda_j$, which are the descending-order eigenvalues of the generally mixed output state $\rho'$.

\begin{figure}[H]%                   FIGURE 3
\centering
\includegraphics[width=0.99\linewidth]{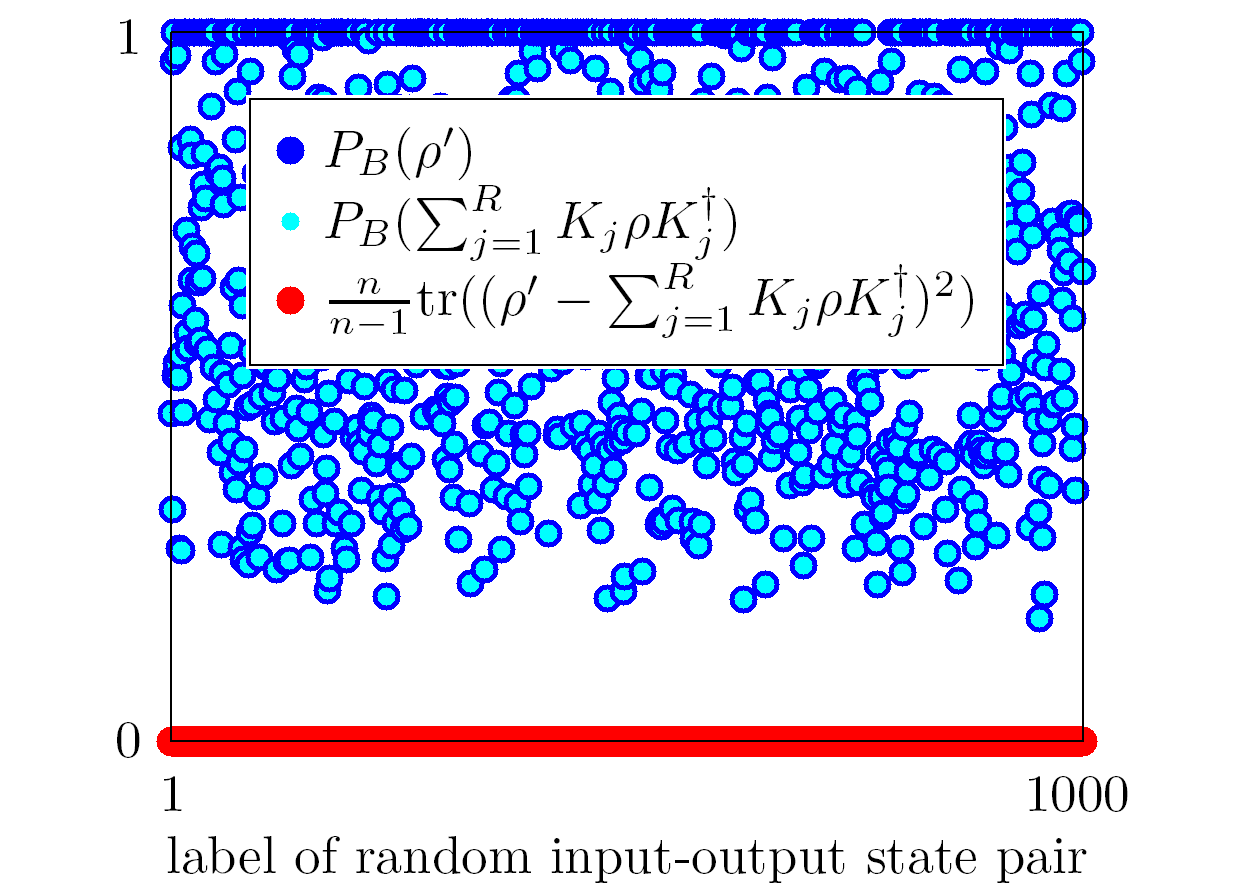}%Base name of figure file here.
\caption[]{\label{fig:3}(color online) Plot of $1000$ random two-qubit input-output state pairs represented two ways each.  First, for visual evidence that different states are used, the height of the dark blue dots is the Bloch purity of the arbitrarily chosen output state $P_{B}(\rho')$, as defined in \Fig{1}.  The height of the light blue dots is the Bloch purity of the RU-reconstructed output state \smash{$P_{B}(\sum_{j=1}^{R}K_{j}\rho K_{j}^{\dag})$}, where the construction is based on the state-dependent Kraus operators of (\ref{eq:83}), and $\rho$ is a randomly chosen pure input state.  However, the \textit{true} necessary and sufficient test is the height of the red dots, which is the square magnitude of the difference of the Bloch vectors of the correct output state and the reconstructed output state, computed as \smash{$\frac{n}{n-1}\text{tr}((\rho'-\sum_{j=1}^{R}K_{j}\rho K_{j}^{\dag})^2)$}.  The reconstruction is successful iff the red dot has a height of zero.  Thus, this method works for all states tested.  Note that it is not limited to two qubits; that is merely an arbitrary choice here.}
\end{figure}

As a demonstration that this works, \Fig{3} plots 1000 random input-output state pairs for two qubits.
%_______________________________________________________________________________
%_______________________________________________________________________________
Note that the restriction to pure input states is not limiting, because it is always possible to purify a mixed input, thus enabling this method.

Again, the \textit{application} of this method is not necessarily useful for general error correction because it requires us to know the input and output states.  

However, suppose that we had some device that produced a particular desired quantum state, but that due to internal imperfections it produces a somewhat corrupted version of the state.  For example, this device could be a laser, inside of which we are certain that it produces an excellent approximation to a pure coherent state, yet random phase kicks cause dephasing.  Since this dephasing is generally time-dependent, that means that we know the output state of this dephasing channel at any time, and of course the input is a pure coherent state.  

Thus, in principle, one could reverse the dephasing effects of a laser field by using the input and output states to form its state-dependent RU-decomposition.

However, the greater worth of the state-dependent RU decomposition is its theoretical suggestion that, in principle, all quantum channels are correctable.  This lends further support to the proof we gave of this using state-\textit{independent} decompositions in \Sec{III.B}.
%                                 END of IV.C
%-------------------------------------------------------------------------------
%                                  END of IV
%*******************************************************************************
%*******************************************************************************
%                              V. CONCLUSIONS
\section{\label{sec:V}Conclusions}
The main goal of this paper is the proof of the correctability of all quantum channels.  To accomplish this, \Sec{II.A} gave an organic, recursive method to obtain an RU decomposition for the $n$-level maximal dephasing channel $\Delta$ and the depolarization channel $\mathcal{D}$ and its extreme form, the maximal-mixing channel $\mathcal{M}$.  This established the fact that RU decompositions exist for the depolarization channel in all dimensions $n$.

Then, in \Sec{II.B}, we developed a method for explicit construction of the RU Kraus operators \smash{$\{M_{(m,\mathbf{x})}\}$} of (\ref{eq:27}) for the maximal mixing channel $\mathcal{M}$, used instead of the depolarization channel because it is more symmetrical in form.  The explicit RU Kraus operators then allowed us to prove that this set always contains at least enough operators to form linear combinations of all the elementary matrices \smash{$E_{(a,b)}^{[n]}$}, which are Hilbert-Schmidt (HS) complete on the set of all matrices, meaning that any matrix can be expressed as a linear combination of the \smash{$E_{(a,b)}^{[n]}$}.  The invertability of the RU Kraus set \smash{$\{M_{(m,\mathbf{x})}\}$} with \smash{$\{E_{(a,b)}^{[n]}\}$} then guarantees that \smash{$\{M_{(m,\mathbf{x})}\}$} can be used to expand any other operator, and is thus HS complete.

In \Sec{III.A}, we reviewed various facts about quantum error-correction codes, including the correctability conditions, and in particular, \Sec{III.A.3} stated that any set of Kraus operators expandable as a linear combination of correctable Kraus operators is also correctable on the same code.  This fact is the basis for the universal correction proof because the HS completeness of the RU Kraus set \smash{$\{M_{(m,\mathbf{x})}\}$} for the maximal mixing channel means that all other sets of Kraus operators can be expanded with \smash{$\{M_{(m,\mathbf{x})}\}$}.  Then, by virtue of the correctability of \smash{$\{M_{(m,\mathbf{x})}\}$} and \Sec{III.A.3}, this means that all quantum channels are correctable with the recovery operation $\mathcal{R}$ that corrects states \smash{$P\rho_{C} P^{\dag}$} in the code space for which the maximal mixing channel is corrected.

We then proved this claim in \Sec{III.B} by first showing that an arbitrary quantum channel $\mathcal{F}$ can have its Kraus operators $F_j$ expanded in terms of a correctable set $\{E_k\}$, which is possible if $\{E_k\}$ is also HS complete.  The proof culminates by showing that the arbitrary channel $\mathcal{F}$ has a Kraus expansion with operators \smash{$\widetilde{F}_r$} which satisfy the correctability conditions for the code with projector $P$ for which $\{E_k\}$ is also correctable.  Thus, arbitrary channels can be corrected, given the existence of an HS-complete correctable set of Kraus operators.

The fact that we had already found an $n$-level set of HS-complete correctable Kraus operators \smash{$\{M_{(m,\mathbf{x})}\}$} proved the existence of the set \smash{$\{E_k\}$} used in the proof for correctability of all quantum channels $\mathcal{F}$, which proved the central claim of this paper.  We then demonstrated that the recovery operation $\mathcal{R}$ does in fact work, restoring states in the code space \smash{$P\rho_{C} P^{\dag}$} up to a proportionality factor, just as a recovery operation should.

The remainder of the paper focused on applications.  First, we used the newly established requirements of HS completeness and correctability to establish a set of conditions for constructing universal recovery operations $\mathcal{R}$ for a code with projector $P$.  These conditions can be used as general guidelines for constructing quantum error correction codes capable of protecting any state against any errors.  Then we showed a specific example of how to verify that any given set of errors is correctable.

Finally, as food for thought, we considered a special topic that showed how to obtain state-dependent RU decompositions of all quantum channels.  While this state-dependence limits its application for general error-correction, we showed that there do exist cases such as the state of a laser field that would benefit from such a state-dependent error-correction scheme.  However, more importantly, the fact that this proves that RU decompositions exist for all possible mixed output states resulting from all possible quantum operations acting on all possible pure input states means that in principle, RU decompositions \textit{always} exist for all quantum channels, even though they may generally need to be state-dependent.  Since RU-decomposable channels are always correctable, the fact that RU decompositions can always be found suggests that all channels \textit{should} be correctable, which further supports the earlier proof of this fact in \Sec{III.B} for state-\textit{independent} methods.

Thus, we have proven that all discrete-system quantum channels are correctable with perfect success.  There are undoubtedly many practical difficulties associated with implementation of the general correction schemes needed to realize such perfect channel correction, however this newly proven fact that it is \textit{always possible} to do so is enormously encouraging news for the field of quantum computation.  It also has profound implications for physics in general because it shows that at least for discrete systems, there is no quantum operation such that its effects cannot be reversed completely with perfect success.  Thus, at least in some local subsystem of interest, it is always possible to reverse any quantum process, even if the corrupted version of the encoded state happens to be the maximally mixed state.

It is likely that there are a large number of applications and explorations possible as a result of these findings, and it is hoped that this paper will provide a useful starting point for further research in quantum error correction and other fields, as well.
%                                  END of V
%*******************************************************************************
%*******************************************************************************
%                               ACKNOWLEDGEMENTS
\begin{acknowledgments}
Many thanks to Ting Yu for his insightful comments.  This project was supported by the I{\&}E Fellowship at Stevens Institute of Technology.
\end{acknowledgments}
%                            END of ACKNOWLEDGEMENTS
%*******************************************************************************
%*******************************************************************************
%                                 BIBLIOGRAPHY
%
%                             END of BIBLIOGRAPHY
%*******************************************************************************
\end{document}